\newcommand*{\ATLASLATEXPATH}{latex/}
\documentclass[cernpreprint,UKenglish,texlive=2011]{\ATLASLATEXPATH atlasdoc}

\usepackage{\ATLASLATEXPATH atlaspackage}
\usepackage{\ATLASLATEXPATH atlasbiblatex}
\usepackage{\ATLASLATEXPATH atlascontribute}
\usepackage{\ATLASLATEXPATH atlasphysics}

\graphicspath{{logos/}{figures/}}

\usepackage{HWWDiffXsecPaper-defs}
\usepackage{multirow}

\AtlasTitle{\boldmath Measurement of fiducial differential cross sections of gluon-fusion production of Higgs bosons decaying to $WW^{\ast}{\rightarrow\,}e\nu\mu\nu$ with the ATLAS detector at $\sqrt{s}=8 \TeV$}

\author{The ATLAS Collaboration}

\AtlasRefCode{HIGG-2015-04}

\PreprintIdNumber{CERN-EP-2016-019}

 \AtlasJournal{J.\ High\ Energy\ Phys.}
 \AtlasJournalRef{JHEP 08 (2016) 104}
 \AtlasDOI{10.1007/JHEP08(2016)104}

\AtlasAbstract{%
  This paper describes a measurement of fiducial and differential cross sections of 
  gluon-fusion Higgs boson production in the $H{\rightarrow\,}WW^{\ast}{\rightarrow\,}e\nu\mu\nu$ channel, using $20.3\,\ifb$ of proton--proton 
collision data. The data were produced at a centre-of-mass energy of $\sqrt{s} = 8 \tev$ at the CERN Large Hadron Collider and recorded by the ATLAS detector in 2012.
  Cross sections are measured from the observed $H{\rightarrow\,}WW^{\ast}{\rightarrow\,}e\nu\mu\nu$ signal yield in categories 
  distinguished by the number of associated jets. 
  The total cross section is measured in a fiducial region defined by the kinematic properties of the charged leptons and neutrinos.
  Differential cross sections are reported as a function of
  the number of jets, the Higgs boson transverse momentum, the dilepton rapidity, and the 
  transverse momentum of the leading jet.  The jet-veto efficiency, or fraction
  of events with no jets above a given transverse momentum threshold, is also reported.
  All measurements are compared 
  to QCD predictions from Monte Carlo generators and fixed-order calculations, and are in agreement with 
  the Standard Model predictions.
}

\AtlasCoverSupportingNote{Differential cross section support note}{https://cds.cern.ch/record/2025025}
\AtlasCoverCommentsDeadline{XX March 2016}

\AtlasCoverAnalysisTeam{Olivier Arnaez, Kathrin Becker, Magda Chelstowska, Biagio Di Micco, Pamela Ferrari, Paul Glaysher, Justin Griffiths, David Hall, Chris Hays, Corrinne Mills, Li Yuan, Li Zhou}

\AtlasCoverEdBoardMember{Fabio Cerutti~(chair), Mike Hance, Joey Huston}

\AtlasCoverEgroupEditors{atlas-higg-2015-04-contact-editors@cern.ch}

\AtlasCoverEgroupEdBoard{atlas-higg-2015-04-editorial-board@cern.ch}

\hypersetup{pdftitle={ATLAS draft},pdfauthor={The ATLAS Collaboration}}

\begin{document}

\maketitle

\tableofcontents

\clearpage

\section{Introduction}
\label{sec:intro}
Since the observation of a new particle by the ATLAS~\cite{HIGG-2012-27} and CMS~\cite{CMS-HIG-12-028} 
collaborations in the search for the Standard Model (SM) Higgs boson~\cite{Englert:1964et,Higgs:1964ia,Higgs:1964pj,Guralnik:1964eu,Higgs:1966ev,Kibble:1967sv},
the mass, spin, and charge conjugation times parity
of the new particle have been measured by both collaborations~\cite{HIGG-2014-14,HIGG-2013-17,CMS-HIG-14-018}. 
Its mass has been measured to be $m_{H}= 125.09 \pm 0.24 \gev$~\cite{HIGG-2014-14} by combining ATLAS and CMS measurements.
The strengths of its couplings to gauge bosons and fermions have also been explored~\cite{HIGG-2014-06,CMS-HIG-14-009}. 
In all cases the results are consistent with SM predictions.
Differential cross-section measurements have recently been made by the ATLAS and CMS collaborations in the $ZZ\to 4\ell$~\cite{HIGG-2013-22,Khachatryan:2015yvw} 
and $\gamma\gamma$~\cite{HIGG-2013-10,CMS-HIG-14-016} final states. 
The results of the ATLAS collaboration have been combined in Ref.~\cite{HIGG-2014-11}.

In this paper, measurements of fiducial and differential cross sections 
for Higgs boson production in the \hwwenmn\  final state are presented. These measurements use $20.3\,\ifb$ of proton--proton 
collision data at a centre-of-mass energy of $\rts = 8 \tev$ recorded by the ATLAS experiment at the CERN
Large Hadron Collider~(LHC). The presented measurements characterise the gluon-fusion production mode~(ggF), which is the 
dominant signal contribution to the \hwwenmn\ event sample.  
The results are compared to quantum chromodynamics~(QCD) predictions of this production mechanism. Small contributions from the 
vector-boson fusion~(VBF), and vector-boson associated production ($VH$) modes are subtracted 
assuming the SM expectation. Contributions from associated Higgs boson production via $t\bar{t}H$ and $b\bar{b}H$ are expected to be negligible 
after applying the experimental event-selection criteria.
To minimise the model dependencies of the correction for the detector acceptance, and to allow direct comparison with theoretical 
predictions, all cross sections presented in this paper are fiducial cross sections corrected for detector effects. Here, the cross sections
are given in a fiducial region defined using particle-level objects where most of the event-selection requirements of the analysis are applied.

The differential ggF Higgs boson production cross sections are chosen to probe several different physical effects:
\begin{itemize}
\item Higher-order perturbative QCD contributions to the ggF production are probed by measuring the number of jets, \Njet, and transverse momentum, $\pT$, of the highest-$\pT$ (``leading'') jet, $\pTj$.
\item Multiple soft-gluon emission, as modelled by resummation calculations, and non-perturbative effects are probed by measuring the transverse momentum of the reconstructed Higgs boson, $\pTH$.
\item Parton distribution functions~(PDFs) are probed by measuring the absolute value of the rapidity of the reconstructed dilepton system, $\yll$.
\end{itemize}
The dilepton rapidity, $y_{\ell\ell}$, is highly correlated to the rapidity of the reconstructed Higgs boson, $y_{H}$, which is known to be sensitive to PDFs. 
Since it is not possible to reconstruct $y_{H}$ experimentally in the \hwwenmn\  final state, the differential cross section is measured as a function of $\yll$.
An additional important test of QCD predictions is the production cross section of the Higgs boson without additional jets ($H+0$-jet),
which is also a significant source of uncertainty in measurements of the total $H\rightarrow WW^{\ast}$ 
production rate. Large uncertainties arise from unresummed logarithms in fixed-order predictions or from uncertainties assigned to resummed predictions for the $H+0$-jet cross section.
The $H+0$-jet cross section, $\sigma_{0}(\pT^{\mathrm{thresh}})$, can be calculated from the product
of the total cross section, $\sigma_{\mathrm{tot}}$, and the jet-veto efficiency for $H+0$-jet events, $\varepsilon_{0}(\pT^{\mathrm{thresh}})$, 
which is defined as the fraction of events with the leading jet below a given threshold, $\pT^{\mathrm{thresh}}$:
\begin{equation}
\sigma_{0}(\pT^{\mathrm{thresh}}) = \varepsilon_{0}(\pT^{\mathrm{thresh}}) \cdot \sigma_{\mathrm{tot}}.
\end{equation} 
In addition to the measurement of the $\Njet$ distribution, a measurement of the jet-veto efficiency for $H+0$-jet events, $\varepsilon_{0}$, is presented 
for three different values of $\pT^{\mathrm{thresh}}$. 
All results are compared to a set of predictions from fixed-order calculations and Monte Carlo~(MC) generators.

Differential cross-section measurements are performed for the first time in the \hwwenmn\ final state. 
This analysis is an extension of the ggF coupling measurement performed using the Run-1 dataset~\cite{HIGG-2013-13},
and uses the same object definitions, background-estimation techniques, and strategies to
evaluate the systematic uncertainties. 
In contrast to the couplings measurement, in which the results were obtained using a likelihood-based approach to simultaneously
fit several signal regions and background-dominated control regions, the analysis
presented here utilizes a simplified approach. First the dominant backgrounds are estimated using control regions in data, and then the predicted backgrounds are subtracted 
from the observed data in the signal region to obtain the signal yield.
Another difference is that events with two leptons of the same flavour ($ee/\mu\mu$) are not considered 
due to the large Drell--Yan ($pp\to{Z/\gamma^{*}}\to\ell\ell$) background.
Using an iterative Bayesian method, the distributions are corrected for detector efficiencies and resolutions.
Statistical and systematic uncertainties are propagated through these corrections, taking correlations among bins into account.

\section{The ATLAS detector}
\label{sec:detector}

The ATLAS detector~\cite{PERF-2007-01} at the LHC covers nearly the entire solid angle around the collision point.
It consists of an inner tracking detector surrounded by a thin superconducting solenoid, electromagnetic and hadronic calorimeters,
and a muon spectrometer incorporating three large superconducting toroid magnets.
The inner-detector system (ID) is immersed in a \SI{2}{\tesla} axial magnetic field 
and provides charged-particle tracking in the range $|\eta| < 2.5$.\footnote{
ATLAS uses a right-handed coordinate system with its origin at the nominal interaction point (IP)
in the centre of the detector and the $z$-axis along the beam pipe.
The $x$-axis points from the IP to the centre of the LHC ring,
and the $y$-axis points upwards.  Cylindrical coordinates $(r,\phi)$ are used in the transverse plane, 
$\phi$ being the azimuthal angle around the $z$-axis.
The pseudorapidity is defined in terms of the polar angle $\theta$ as $\eta = -\ln \tan(\theta/2)$.
Angular separation is measured in units of $\Delta R \equiv \sqrt{(\Delta\eta)^{2} + (\Delta\phi)^{2}}$.}

Closest to the interaction point, the silicon-pixel detector forms the three innermost layers of the inner detector.  
The silicon-microstrip tracker surrounding it typically provides four additional two-dimensional measurement points 
per track.  The silicon detectors are complemented by the transition-radiation tracker,
which enables radially extended track reconstruction up to $|\eta| = 2.0$ and  
provides electron identification information based on the fraction of hits above a higher energy-deposit threshold indicating
the presence of transition radiation.

The calorimeter system covers the range $|\eta| < 4.9$.
Within the region $|\eta|< 3.2$, electromagnetic calorimetry is provided by
a high-granularity lead/liquid-argon (LAr) sampling calorimeter.
The hadronic calorimeter consists of steel and scintillator tiles in the central region and
two copper/LAr hadronic endcap calorimeters.
The solid-angle coverage is completed with forward copper/LAr and tungsten/LAr calorimeter modules
optimised for electromagnetic and hadronic measurements respectively.

The muon spectrometer (MS) covers the region $|\eta| < 2.7$ with precise position measurements from three layers 
of monitored drift tubes (MDTs).  Cathode-strip chambers provide additional high-granularity coverage in the forward ($2 < |\eta| < 2.7$) region.
The muon trigger system covers the range $|\eta| < 2.4$ with resistive-plate chambers in the barrel and thin-gap chambers 
in the endcap regions, both of which also provide position measurements in the direction normal to the bending plane, 
complementary to the precision hits from the MDTs.  

A three-level trigger system reduces the event rate to about \SI{400}{\Hz}~\cite{PERF-2011-02}.
The Level-1 trigger is implemented in hardware and uses a subset of detector information
to reduce the event rate to a design value of at most \SI{75}{\kHz}.
The two subsequent trigger levels, collectively referred to as the High-Level Trigger (HLT), are implemented in software.

\section{Signal and background models}
\label{sec:mcpred}
Signal and background processes are modelled by Monte Carlo simulation, using the same samples
 and configurations as in Ref.~\cite{HIGG-2013-13}, which are summarized here.
Events representing the ggF and VBF $H\to{WW^{\ast}}$ signal processes are produced from calculations at next-to-leading 
order~(NLO) in the strong coupling $\alphas$ as implemented in the
\POWHEG\ MC generator~\cite{Nason:2004rx,Alioli:2008tz,Nason:2009ai,Bagnaschi:2011tu}, interfaced with \PythiaEight \cite{Sjostrand:2007gs} (version 8.165) 
for the parton shower, hadronisation, and underlying event.  The CT10~\cite{Lai:2010vv} PDF set is used and the parameters of the \PythiaEight generator controlling the modelling of the parton shower
and the underlying event are those corresponding to the AU2 set~\cite{ATL-PHYS-PUB-2012-003}. 
The Higgs boson mass set in the generation is 125.0 \GeV, which
is close to the measured value. 
The {\POWHEG} ggF model takes into account finite quark masses and a running-width Breit--Wigner
distribution that includes electroweak corrections at NLO~\cite{deFlorian:2012yg}.  
To improve the modelling of the Higgs boson $\pT$ distribution, a reweighting scheme is applied to reproduce 
the prediction of the next-to-next-to-leading-order~(NNLO) and next-to-next-to-leading-logarithm~(NNLL) dynamic-scale calculation given by the \HRES2.1 program~\cite{Grazzini:2013mca}.  
Events with ${\geq\,}2$ jets are further reweighted to reproduce the $\pTH$ spectrum predicted by the NLO 
{\POWHEG} simulation of Higgs boson production in association with two jets ($H\,+\,2$~jets)~\cite{Hamilton:2012np}.  
Interference with continuum $WW$ production~\cite{Campbell:2011cu, Campbell:2013wga} has 
a negligible impact on this analysis due to the transverse-mass selection criteria described in Section~\ref{sec:selection} and is not 
included in the signal model.

The inclusive cross sections at $\rts = 8 \tev$ for a Higgs boson mass of $125.0 \gev$, calculated at NNLO+NNLL in QCD and NLO in the electroweak couplings, 
are 19.3 pb and 1.58 pb for ggF and VBF respectively~\cite{YR3}. The uncertainty on the ggF cross
section has approximately equal contributions from QCD scale variations (7.5\%) and PDFs (7.2\%). For the VBF production, the uncertainty
on the cross section is 2.7\%, mainly from PDF variations.
The $WH$ and $ZH$ processes are modelled with \PythiaEight and normalised 
to cross sections of 0.70 pb and 0.42 pb respectively, calculated at NNLO in QCD and NLO in the electroweak couplings~\cite{YR3}.  
The uncertainty is 2.5\% on the $WH$ cross section and 4.0\% on the $ZH$ cross section.

For all of the background processes, with the exception of $\Wjets$ and multijet events, MC simulation is used to model 
event kinematics and as an input to the background normalisation.  The $\Wjets$ and multijet background models are
derived from data as described in Section~\ref{sec:background}.
For the dominant $WW$ and top-quark backgrounds, the MC generator is \POWHEG+\PythiaSix~\cite{Sjostrand:2006za} (version 6.426),
also with CT10 for the input PDFs. The Perugia 2011 parameter set is used for \PythiaSix~\cite{Skands:2010ak}. 
For the $WW$ background with $\TwoJet$, to better model the additional partons, the \SHERPA~\cite{Gleisberg:2008ta} program 
(version 1.4.3) with the CT10 PDF set is used.  The Drell--Yan background, 
including $\Ztt$, is simulated with the \ALPGEN~\cite{Mangano:2006rw} program (version 2.14). 
It is interfaced with \HERWIG~\cite{Corcella:2000bw} (version 6.520) with parameters set to those of the ATLAS Underlying Event Tune 2~\cite{ATL-PHYS-PUB-2011-008} and uses the CTEQ6L1~\cite{Nadolsky:2008zw} PDF set.  
The same configuration is applied for $W\gamma$ events.  Events in the $Z/\gamma^{\ast}$ sample are reweighted to 
the MRSTmcal PDF set~\cite{Sherstnev:2007nd}.  For the $W\gamma^{\ast}$ and $Z/\gamma$ backgrounds, the \SHERPA\ program is used, with
the same version number and PDF set as the $WW$ background with $\geq\,2$ jets.  
Additional diboson backgrounds, from $WZ$ and $ZZ$, are modelled using
\POWHEG+\PythiaEight.  

For all MC samples, the ATLAS detector response is simulated~\cite{SOFT-2010-01} using 
either \GeantFour~\cite{GEANT4} or \GeantFour combined with a parameterised \GeantFour-based calorimeter 
simulation~\cite{ATL-PHYS-PUB-2010-013}.  Multiple proton--proton~(pile-up) interactions are modelled by overlaying minimum-bias
interactions generated using \PythiaEight.

\section{Event selection}
\label{sec:selection}
This section describes the reconstruction-level definition of the signal region.
The definition of physics objects reconstructed in the detector follows that of Ref.~\cite{HIGG-2013-13} 
exactly and is summarised here.  
All objects are defined with respect to a primary interaction vertex, which is required to
have at least three associated tracks with $\pT\,\ge\,400\MeV$.  If more than one such vertex is
present, the one with the largest value of $\sum(\pT^2)$, where the sum is over all tracks associated with that vertex, is selected as the primary vertex.

\subsection{Object reconstruction and identification}

Electron candidates are built from clusters of energy depositions in the 
EM calorimeter with an associated well-reconstructed track.  They are required to 
have $\ET>10\GeV$, where the transverse energy $\ET$ is defined as $E\sin(\theta)$.  Electrons reconstructed
with $|\,\eta\,|\,{<}\,2.47$ are used, excluding $1.37\,{<}\,|\,\eta\,|\,{<}\,1.52$, which corresponds to the 
transition region between the barrel and the endcap calorimeters.
Additional identification criteria are applied to reject background, using the 
calorimeter shower shape, the quality of the match between the track and the 
cluster, and the amount of transition radiation emitted in the ID~\cite{PERF-2013-03,ATLAS-CONF-2014-032,PERF-2013-05}.
For electrons with $10\GeV < \ET < 25\GeV$, a likelihood-based
electron selection at the ``very tight'' operating point is used
for its improved background rejection.  For $\ET > 25\GeV$, 
a more efficient ``medium'' selection is used because background is less of a concern.
The efficiency of these requirements varies strongly as a function of $\ET$, 
starting from 65--70\% for $\ET < 25\GeV$, jumping to about 80\% with the change in 
identification criteria at $\ET = 25\GeV$, and then steadily increasing as a function of $\ET$~\cite{ATLAS-CONF-2014-032}.

Muon candidates are selected from tracks reconstructed in the ID matched to tracks 
reconstructed in the muon spectrometer. Tracks in
both detectors are required to have a minimum number of hits to ensure robust reconstruction.
Muons are required to have $|\,\eta\,|\,{<}\,2.5$ and $\pT>10\GeV$.  
The reconstruction efficiency is between 96\% and 98\%, and stable as a function of $\pT$~\cite{PERF-2014-05}.

Additional criteria are applied to electrons and muons to reduce 
backgrounds from non-prompt leptons and electromagnetic signatures produced by hadronic activity. 
Lepton isolation is defined using track-based and calorimeter-based quantities. 
All isolation variables used are normalised relative to the transverse momentum of the lepton,
and are optimised for the $\hwwenmn$ analysis, resulting in stricter criteria for
better background rejection at lower $\pT$ and looser criteria for better efficiency
at higher $\pT$.  Similarly, requirements on the transverse impact-parameter 
significance $d_0/\sigma_{d_0}$ and the longitudinal impact parameter $z_0$ are made.
The efficiency of the isolation and impact-parameter requirements for electrons satisfying all of the identification 
criteria requirements ranges from 68\% for $10\GeV < \ET < 15\GeV$ to greater than 90\% for 
electrons with $\ET > 25\GeV$.  For muons, the equivalent efficiencies are 60--96\%.

Jets are reconstructed from topological clusters of calorimeter
cells~\cite{PERF-2012-01,ATL-LARG-PUB-2008-002,ATLAS-CONF-2014-018} using
the anti-$k_{t}$ algorithm with a radius parameter of $R=0.4$~\cite{Cacciari:2008gp}. 
Jet energies are corrected for the effects of calorimeter non-compensation,
signal losses due to noise threshold effects, energy lost in non-instrumented regions,
contributions from in-time and out-of-time pile-up, and the position of the primary interaction 
vertex~\cite{PERF-2012-01,ATLAS-CONF-2012-064}.
Subsequently, the jets are calibrated to the hadronic energy scale~\cite{PERF-2011-03,PERF-2012-01}.
To reduce the chance of using a jet produced by a pile-up interaction, jets with with $\pT < 50 \gev$ 
and $|\eta| < 2.4$ are required to have more than 50\% of the scalar sum 
of the \pt\ of their associated tracks come from tracks associated with the primary vertex.  
Jets used for definition of the signal region are required to have 
$\pT\,{>}\,25\GeV$ if $|\,\eta\,|\,{<}\,2.4$ and $\pT\,{>}\,30\GeV$ if
$2.4\,{<}\,|\,\eta\,|\,{<}\,4.5$.  

Jets containing $b$-hadrons are identified using a multivariate
$b$-tagging algorithm~\cite{ATLAS-CONF-2014-004,ATLAS-CONF-2014-046} 
which combines impact-parameter information of tracks and the
reconstruction of charm- and bottom-hadron decays.  The working point, chosen to maximise 
top-quark background rejection, has an efficiency of 85\% for $b$-jets and a
mis-tag rate for light-flavour jets (excluding jets from charm quarks) of 10.3\% in simulated \ttbar\ events.  

Missing transverse momentum ($\vmet$) is produced in signal events by the two 
neutrinos from the $W$ boson decays.  It is reconstructed as the 
negative vector sum of the transverse momenta of muons, electrons, photons, 
jets, and tracks with $\pT > 0.5 \gev$ associated with the primary vertex but unassociated with any of the previous objects.

\subsection{Signal region selection}

Events are selected from those with exactly one electron and one muon with opposite charge, a dilepton invariant 
mass $\mll$ greater than $10\GeV$, and $\met\,{>}\,20\GeV$.  At least one of the two leptons is required to have $\pT\,{>}\,22\GeV$
and the lepton with higher $\pT$ is referred to as the leading lepton.  The other (``subleading'') lepton is required to 
have $\pT\,{>}\,15\GeV$.  All events are required to pass at least one single-lepton or dilepton trigger.  The 
Level-1 $\pT$ thresholds for the single-lepton triggers are $18\GeV$ and $15\GeV$ for electrons and muons, respectively.
The HLT uses object reconstruction and calibrations close to those used offline, and the electron and muon
triggers both have thresholds at $24\GeV$ and an isolation requirement.  To recover efficiency, a supporting trigger 
with no isolation requirement but higher
$\pT$ thresholds, $60\GeV$ for electrons and $36\GeV$ for muons, is used.  The dilepton trigger requires an electron and a muon 
above a threshold of $10\GeV$ and $6\GeV$, respectively, at Level-1, and $12\GeV$ and $8\GeV$ in the HLT.  This increases the signal 
efficiency by including events with a leading lepton
below the threshold imposed by the single-lepton triggers but still on the plateau of the dilepton trigger efficiency.
The reconstructed leptons are required to match those firing the trigger.  
The total per-event trigger efficiencies for events with $\ZeroJet$ are 96\% for events with a leading 
electron and 84\% for events with a leading muon.  The efficiency increases with increasing jet multiplicity,
up to 97\% for events with a leading electron and 89\% for events with a leading muon. 

Three non-overlapping signal regions are defined, distinguished by the number of reconstructed jets: 
$\ZeroJet$, $\OneJet$, or $\TwoJet$.  These separate the data into signal regions with different background compositions,
which improves the sensitivity of the analysis.  The dominant background processes are $WW$ production for $\ZeroJet$, 
top-quark production for $\TwoJet$, and a mixture of the two for $\OneJet$.  For jet multiplicities above two, the number of events decreases with 
increasing number of jets but the background composition remains dominated by top-quark production, so these events are all collected in the 
$\TwoJet$ signal region.  

The signal regions are based on the selection used for the ggF analysis of Ref.~\cite{HIGG-2013-13}, with modifications to improve 
the signal-to-background ratio, and to account for the treatment of VBF and $VH$ as backgrounds.  
The former includes the increase in the subleading lepton $\pT$ threshold and the exclusion of same-flavour events, to reduce
background from $\Wjets$ and Drell--Yan events, respectively.  

The selection criteria are summarised in Table~\ref{tab:selection}.  
The $b$-jet veto uses jets with $\pT > 20\GeV$ and $|\eta^{\mathrm{jet}}|\,<\,2.4$, and rejects top-quark background
in the $\OneJet$ and $\TwoJet$ categories.  Background from $\Ztt$ and multijet events is reduced in the $\ZeroJet$ category 
with a requirement on the transverse momentum of the dilepton system, $\pTll\,>\,30\GeV$.  In the $\OneJet$ category, this 
is accomplished in part by requirements on the single-lepton transverse mass $\mT^{\ell}$, defined for each lepton as 
$\mT^{\ell} = \sqrt{2(\met \pT^{\ell}\,-\,\vpTl\cdot\vmet)}$.  At least one of the two leptons is required to have $\mT^{\ell}\,{>}\,50\GeV$.  
For $\Ztt$ background events in the $\OneJet$ and $\TwoJet$ categories, the $\pT$ of the $\tau\tau$ system is larger, so the 
collinear approximation is used to calculate the $\tau\tau$ invariant mass $\mtt$~\cite{Plehn:1999xi}.  A requirement that
$\mtt$ at $m_Z - 25\GeV$ suppresses most background from $\Ztt$.  
Selection that rejects $\Ztt$ events also rejects $H\to\tau\tau$ events, which are kinematically similar.
The VBF veto in the $\TwoJet$ signal region removes events in which the two leading jets have an invariant mass $\mjj\,{>}\,600\GeV$ 
and a rapidity separation $\dyjj\,{>}\,3.6$, which rejects about 40\% of VBF events but only 5\% of ggF events.  

Upper bounds on $\mll$ and the azimuthal angle between the leptons
$\dphill$ take advantage of the unique kinematics of the $H\to{WW^{\ast}}$ decay to discriminate between these 
signal events and the continuum $WW$ background. The 
spin-zero nature of the Higgs boson, together with the structure of the weak interaction in the $W$ boson decays, preferentially produces leptons
pointing into the same hemisphere of the detector.  The small
dilepton invariant mass is a consequence of that and the fact that $m_H\,<\,2m_W$, which forces one of the two $W$ bosons off-shell,
resulting in lower lepton momenta in the centre-of-mass frame of the Higgs boson decay.  
\begin{table}[t!]
  \centering
  \caption{
    Event selection criteria used to define the signal regions in the \hwwenmn\ differential cross section measurements. 
      The preselection and signal-topology selection criteria are identical across all signal regions. 
      The background rejection and VBF-veto selection depend on $\Njet$, and a dash (`-') indicates that no
      selection is applied.  Definitions including the $\pT$ thresholds for jet counting are given in the text.
  }
  \begin{tabular}{llll}
  \dbline
  Category	    & $\ZeroJet$ & $\OneJet$ & $\TwoJet$ \\
  \sgline
  Preselection     &
  \multicolumn{3}{c}{
  \begin{tabular}{c}
  Two isolated leptons ($\ell\,{=}\,e, \mu$) with opposite charge\\
  $\pTlead\,{>}\,22\GeV$, $\pTsublead\,{>}\,15\GeV$ \\
  $\mll\,{>}\,10\GeV$ \\
  $\met\,{>}\,20\GeV$	\\
  \end{tabular}
  }\\
  \sgline
  Background rejection	   & - 		& $\Nbjet\,{=}\,0$   & $\Nbjet\,{=}\,0$ \\
		    & $\Delta\phi(\ell\ell,\met) > 1.57$	& max$(\mT^{\ell})\,{>}\,50\GeV$ 	& - 	\\
  		    & $\pTll\,{>}\,30\GeV$ & $\mtt\,{<}\,m_Z - 25\GeV$ & $\mtt\,{<}\,m_Z - 25\GeV$	\\
  \sgline
  VBF veto	& - 		& - 		& $\mjj\,{<}\,600\GeV$ or $\dyjj\,{<}\,3.6$	\\
  \sgline
  \multirow{2}{*}{\!\!\!\!\!
  \begin{tabular}{l}
  $\hwwlnln$ \\
  topology
  \end{tabular}
  }
  		    & \multicolumn{3}{c}{$\mll\,{<}\,55\GeV$}  \\
  		    & \multicolumn{3}{c}{$\dphill\,{<}\,1.8$} \\
  		    & \multicolumn{3}{c}{$85\GeV\,{<}\,\mT\,{<}\,125\GeV$} \\
  \dbline
  \end{tabular}
  \label{tab:selection}
\end{table}

Signal events are peaked in the distribution of the transverse mass \mT, defined as 
\begin{equation}
 \mT = \sqrt{(\ET^{\ell\ell} + \met)^2 - |\vpTll + \vmet|^2} , 
\end{equation}
  where 
\begin{equation}
  \ET^{\ell\ell} = \sqrt{|\vpTll|^2 + m_{\ell\ell}^2} . 
\end{equation}
Figure~\ref{fig:SR_MT} shows the \mT\ distribution after application of all other selection criteria in each of the signal regions.
Selecting events with $85\GeV\,{<}\,\mT\,{<}\,125\GeV$ increases the signal region purity and  
minimises the total uncertainty of this measurement of the ggF cross section.
Removing events with $\mT\,\gtrsim\,m_H$ also reduces the effect of interference with the continuum $WW$ process
to negligible levels compared to the observed event yield~\cite{Campbell:2011cu}.  

The distributions to be measured are built using the same leptons, jets, and $\met$ that enter the event selection.
The $\vpT$ of the Higgs boson (\pTH) is reconstructed as the vector sum of the missing transverse momentum
and the $\vpT$ of the two leptons:
\begin{equation}
  \pTH = |\vpTlead + \vpTsublead + \vmet| . 
\end{equation}
The rapidity of the dilepton system $\yll$ is reconstructed from the charged lepton
four momenta.  The reconstructed and unfolded distributions are binned using the bin edges 
defined in Table~\ref{tab:binedge}.
\begin{table}[tb!]
  \centering
  \caption{ Bin edges for the reconstructed and unfolded distributions.  }
  \begin{tabular}{rl}
  \dbline
$\pTH$ $[\GeV]$: 	& [0--20], [20--60], [60--300]	\\
$\yll$: 	& [0.0--0.6], [0.6--1.2], [1.2--2.5]	\\
$\pTj$ $[\GeV]$:	& [0--30], [30--60], [60--300]	\\
  \dbline
  \end{tabular}
  \label{tab:binedge}
\end{table} 
The bin edges are determined by balancing the expected statistical and systematic uncertainties in each bin. The resolution 
of the variables is smaller than the bin size and does not affect the binning choice.
For each distribution, the upper edge of the highest bin is chosen so that less than 1\% of the expected event yield in the fiducial region is excluded.

\begin{figure}[!bth]
  \centering
  \subfloat[$\ZeroJet$] {
\includegraphics[width=0.46\textwidth]{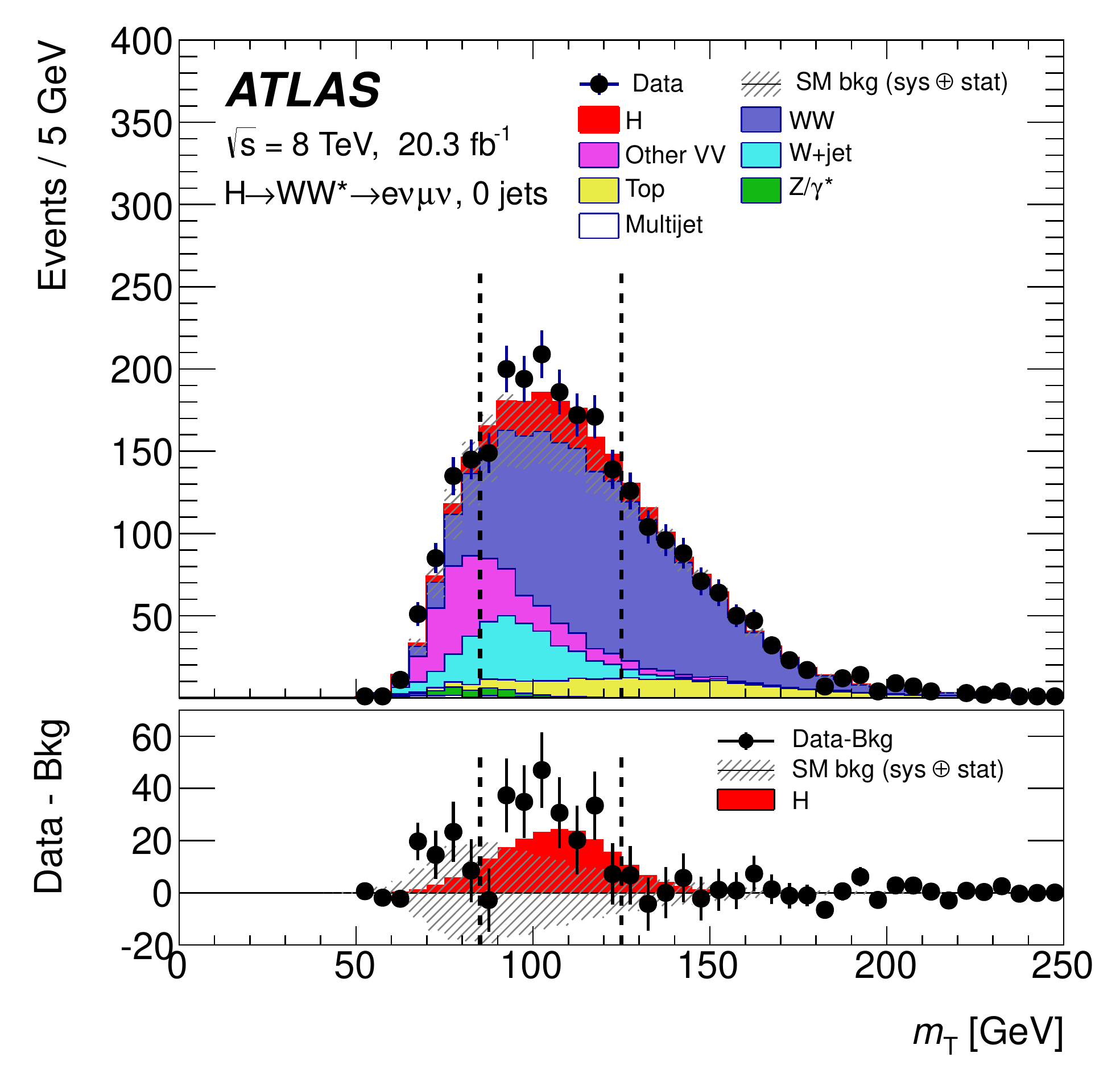}
  }
  \subfloat[$\OneJet$] {
\includegraphics[width=0.46\textwidth]{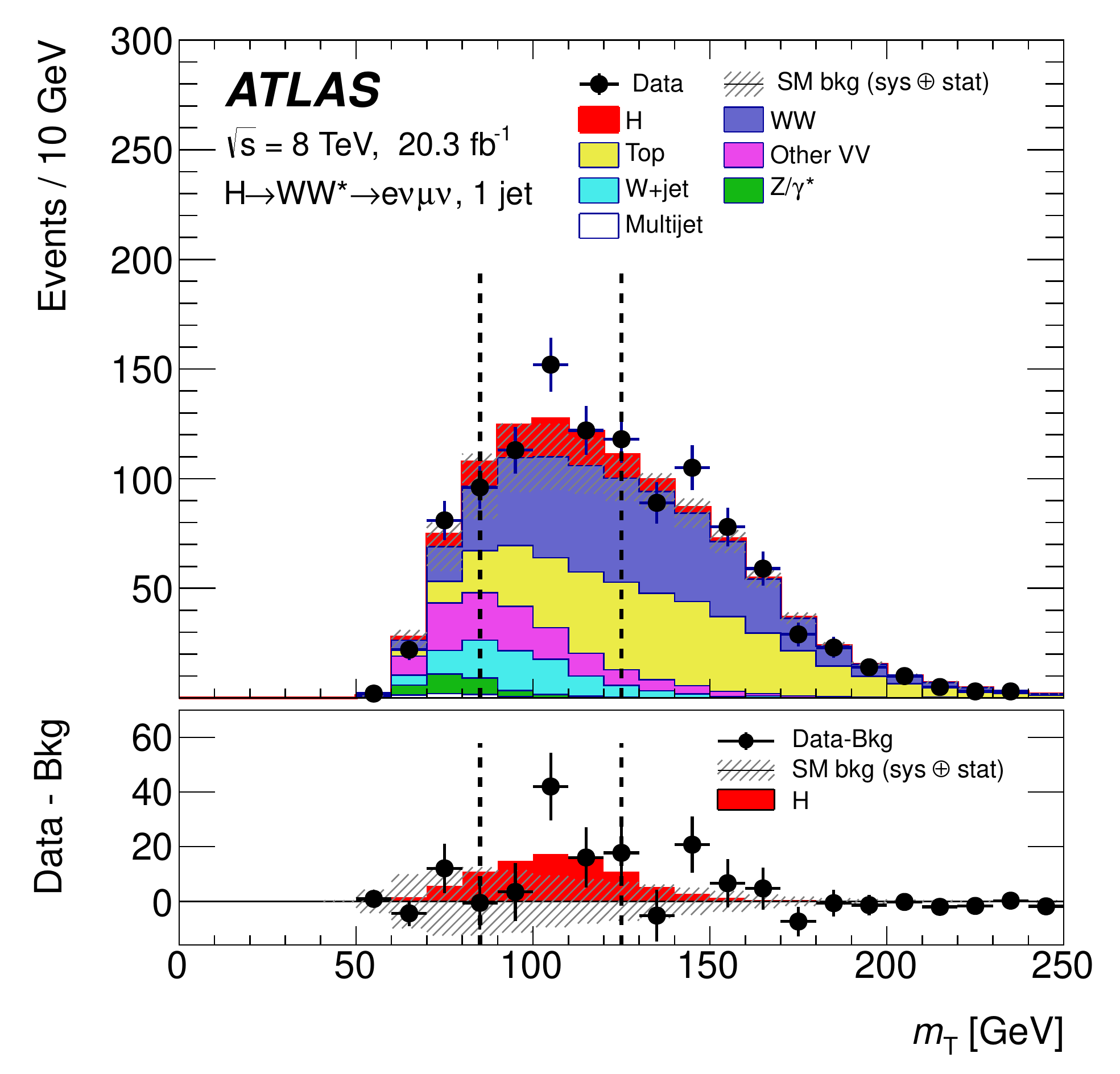}
  }
\newline
  \subfloat[$\TwoJet$] {
\includegraphics[width=0.46\textwidth]{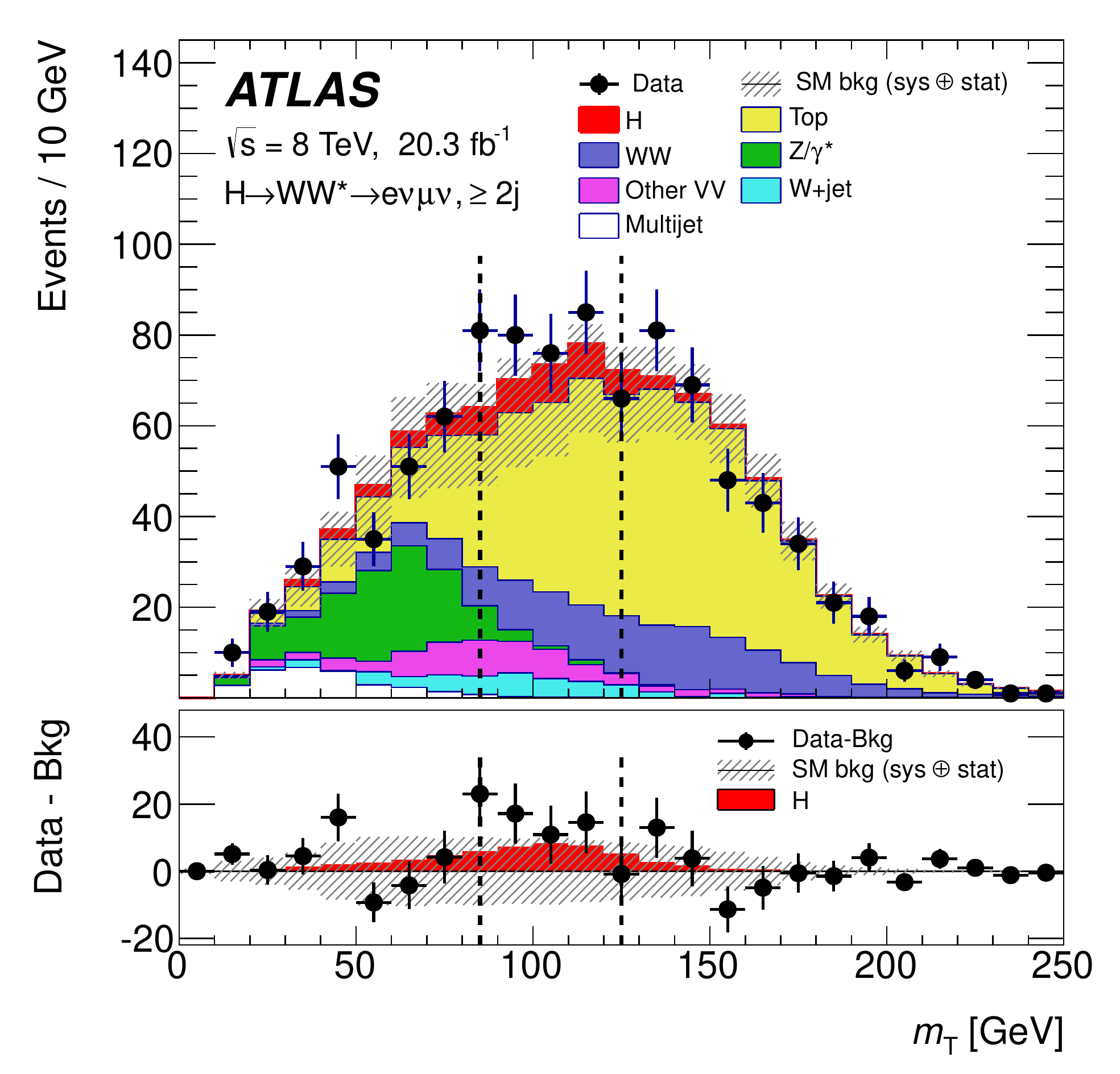}
  }
  { \caption{Observed distributions of \mT\ with signal and background expectations after all other selection criteria have been applied 
	      for the \ZeroJet\ (top left), \OneJet\ (top right) and \TwoJet (bottom) signal regions. 
		The background contributions are normalised as described in Section~\ref{sec:background}.  
		  The SM Higgs boson signal prediction shown is summed over all production processes.
  The hatched band shows the sum in quadrature of statistical and systematic uncertainties of the sum of the backgrounds. 
  The vertical dashed lines indicate the lower and upper selection boundaries on \mT\ at 85 and $125 \GeV$. 
  \label{fig:SR_MT}
    }}
\end{figure}

\section{Background estimation}
\label{sec:background}
Important background processes for this analysis are $WW$, $t\bar{t}$, single top-quark,
$\Ztt$, $\Wjets$, and diboson processes other than $WW$, collectively referred to as ``Other $VV$'' and including
$W\gamma^*$, $W\gamma$, $WZ$, and $ZZ$ events.  The background estimation techniques are described in detail in Ref.~\cite{HIGG-2013-13} and briefly here.  The 
normalisation strategy is summarised in Table~\ref{tab:ggFCR012j}.
As much as possible, backgrounds are estimated using a control region (CR) enriched in the target background
and orthogonal to the signal region (SR), because the statistical and extrapolation uncertainties are smaller than
the typical uncertainties associated with explicit prediction of the yields in exclusive $\Njet$ categories. 
The background estimates done in the CRs are extrapolated to the SR using extrapolation factors taken from simulation.
The control region definitions are summarised in Table~\ref{tab:CRselection}, and 
include the lower subleading lepton $\pT$ threshold of $10\GeV$ for all control regions except the one for $WW$.  This is done  
because the gain in statistical precision of the resulting background estimates is larger than the increase of the systematic uncertainties on the extrapolation factors, particularly for the $\Ztt$ and $VV$ processes.

For all kinematic distributions, except $\Njet$, the shapes are derived from data for 
the $\Wjets$ and multijet backgrounds, and from the MC-simulated background samples for all other processes.
Because the signal regions are defined in terms of $\Njet$, the $\Njet$ distribution is determined directly in each
bin by the sum of the background predictions.
Theoretical and experimental uncertainties are evaluated for all MC-simulation-derived shapes and included in the analysis,
as described in Section~\ref{sec:uncerts}. 

The contribution to the signal region from the VBF and $VH$ Higgs boson production modes, and all contributions from $H\to\tau\tau$ decays,
are treated as a background assuming the Standard Model cross section, branching ratio, and acceptance for $\mH=125\GeV$.   
The contribution of $H\to\tau\tau$ events is negligible due to the selection criteria rejecting $\tau\tau$ events.   
The largest contribution from all non-ggF Higgs boson processes is in the $\TwoJet$ category, in which events from 
VBF and $VH$ contribute about half the number of events that ggF does, and constitute about 3\% of the total background.  
The $\Njet$ distribution and other shapes are taken from simulation.  

\begin{table*}[tb!]
\centering
\caption{
Summary of background-estimation procedures for the three signal regions.  Each background
  is categorised according to whether it is
normalised using a control region (CR), a fully data-derived estimate (Data),
or the theoretical cross section and acceptance from simulation (MC).  
}
\begin{tabular}{llllllll}
\dbline
Channel		& $WW$ & Top 	     & \Ztt\     & \Zeemm & $W+$jets/multijet & Other $VV$ \\
\sgline
\ZeroJet & CR         & CR         & CR    	 & MC    & Data 	&  CR \\
\OneJet & CR         & CR         & CR        & MC    & Data 	&  CR \\
\TwoJet & MC         & CR         & CR        & MC    & Data 	& MC  \\
  \dbline
\end{tabular}
\label{tab:ggFCR012j}
\end{table*}

\begin{table*}[t!]
  \centering
  \caption{
    Event selection criteria used to define the control regions.
    Every control region starts from the same basic charged lepton and $\met$ selection as the signal regions
    except that the subleading lepton $\pT$ threshold is lowered to $10\GeV$ unless otherwise stated.
    Jet-multiplicity requirements also match the corresponding signal region, except where noted for some top-quark control regions.
    The ``top quark aux.'' lines describe auxiliary data control regions used to correct the normalisation
    found in the main control region.  Dashes indicate that a particular control region is not defined. 
    The definitions of $\mtt$, $\mT^{\ell}$, and the jet counting $\pT$ thresholds are as for the signal regions.
  }
  \begin{tabular}{llll}
  \dbline
  CR		& $\ZeroJet$ & $\OneJet$ & $\TwoJet$ \\
  \sgline
  $WW$		& $55\,{<}\,\mll\,{<}\,110\GeV$ & $\mll\,{>}\,80\GeV$		& 	- 	\\
  		& $\dphill\,{<}\,2.6$		& $|\mtt-m_Z|\,{>}\,25\GeV$ 	& 	\\
  		& $\pTsublead\,{>}\,15\GeV$	& $\pTsublead\,{>}\,15\GeV$	& 	\\
		& 				& $b$-jet veto			& 	\\
		& 				& max$(\mT^{\ell})\,{>}\,50\GeV$	& 	\\
\sgline
Top quark	& No $\Njet$ requirement	& $\ge\,1$ $b$-jet required	& $\mll\,{>}\,80\GeV$	\\
      		& $\dphill\,{<}\,2.8$		& 				& $b$-jet veto		\\
\sgline
Top quark aux.	& No $\Njet$ requirement	& $\Njet = 2$			& 		\\
		& $\ge\,1$ $b$-jet required	& $\ge\,1$ $b$-jet required	& -		\\
\sgline
Other $VV$	& Same-sign leptons	& Same-sign leptons			& - \\
    		& All SR cuts		& All SR cuts				& 	\\
\sgline
$\Ztt$		& $\mll < 80\GeV$	& $\mll < 80\GeV$	& $\mll < 70\GeV$	\\
    		& $\dphill > 2.8$	& $\mtt > m_Z - 25\GeV$	& $\dphill > 2.8$	\\
		& 			& $b$-jet veto		& $b$-jet veto		\\
\dbline
  \end{tabular}
  \label{tab:CRselection}
\end{table*} 

For the $\ZeroJet$ and $\OneJet$ categories, the $WW$ background is normalised using control regions 
distinguished from the SR primarily by $\mll$, and the shape is taken from simulated events generated using \POWHEG+\PythiaSix\ as 
described in Section~\ref{sec:mcpred}.  For the $\TwoJet$ category, $WW$ is 
normalised using the NLO cross section calculated with MCFM~\cite{Campbell:2006xx}.  The efficiency for the
$\TwoJet$ requirement and other SR selections is taken from MC simulation, for which  
the $\SHERPA$ generator is used.  It is LO in QCD but has matrix elements implemented
for $WW\,+\,N$ jets, for $0\le N \le 3$.  
For all $\Njet$ categories, $WW\to\ell\nu\ell\nu$ background events produced by double parton scattering are normalised using the predicted 
cross section times branching ratio of $0.44\,\pm\,0.26$ pb~\cite{HIGG-2013-13}.
The acceptance is modelled at LO using events generated by \PythiaEight.   
The $\yll$ distribution in the $\ZeroJet$ $WW$ CR and the $\pTH$ distribution in the $\OneJet$ $WW$ CR are
shown in Figure~\ref{fig:WWCR}.

\begin{figure}[tbp!]
  \centering
  \subfloat[$\yll$, $\ZeroJet$]{
	\includegraphics[width=0.46\textwidth]{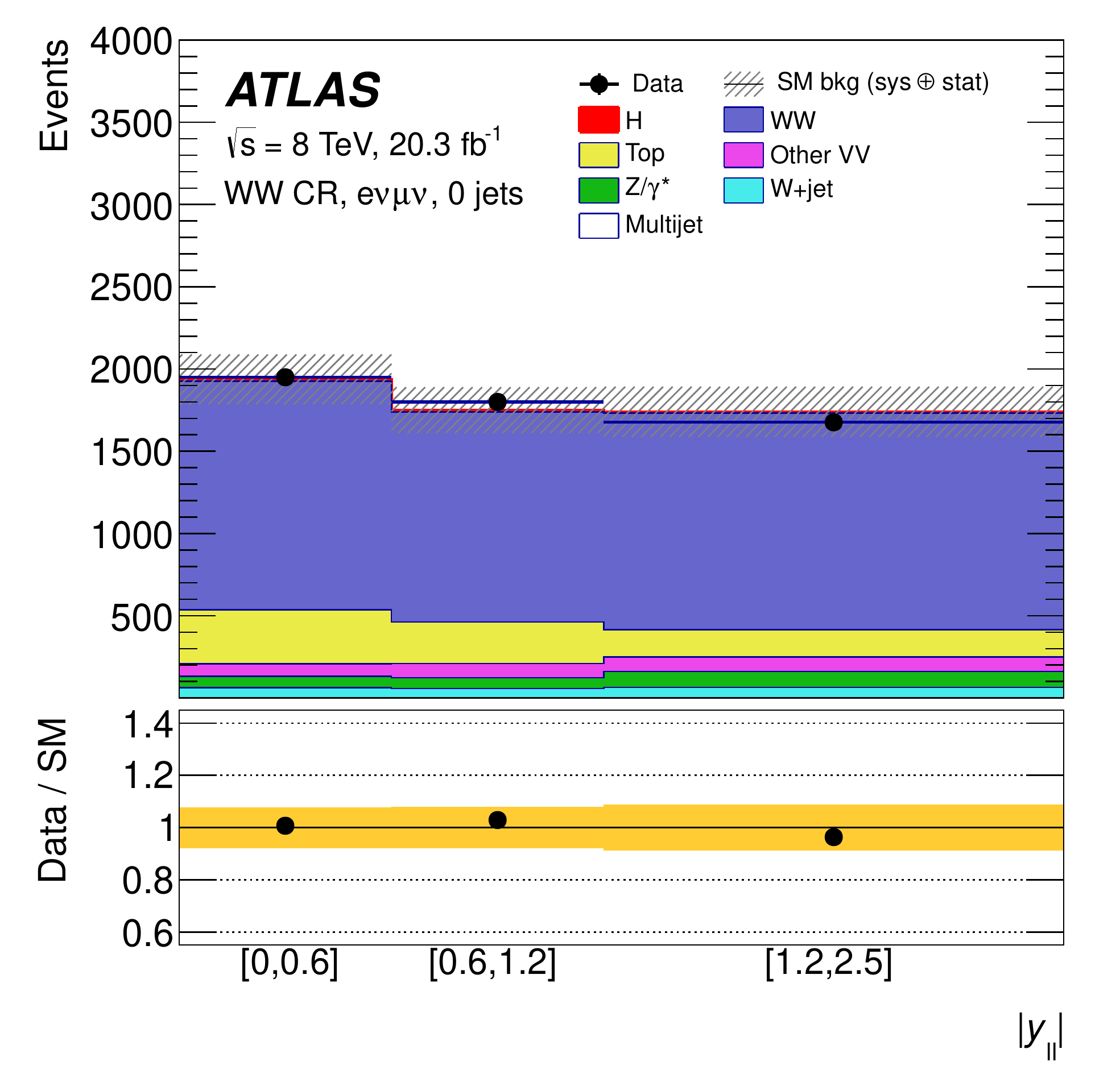}
  }
  \subfloat[$\pTH$, $\OneJet$]{
	\includegraphics[width=0.46\textwidth]{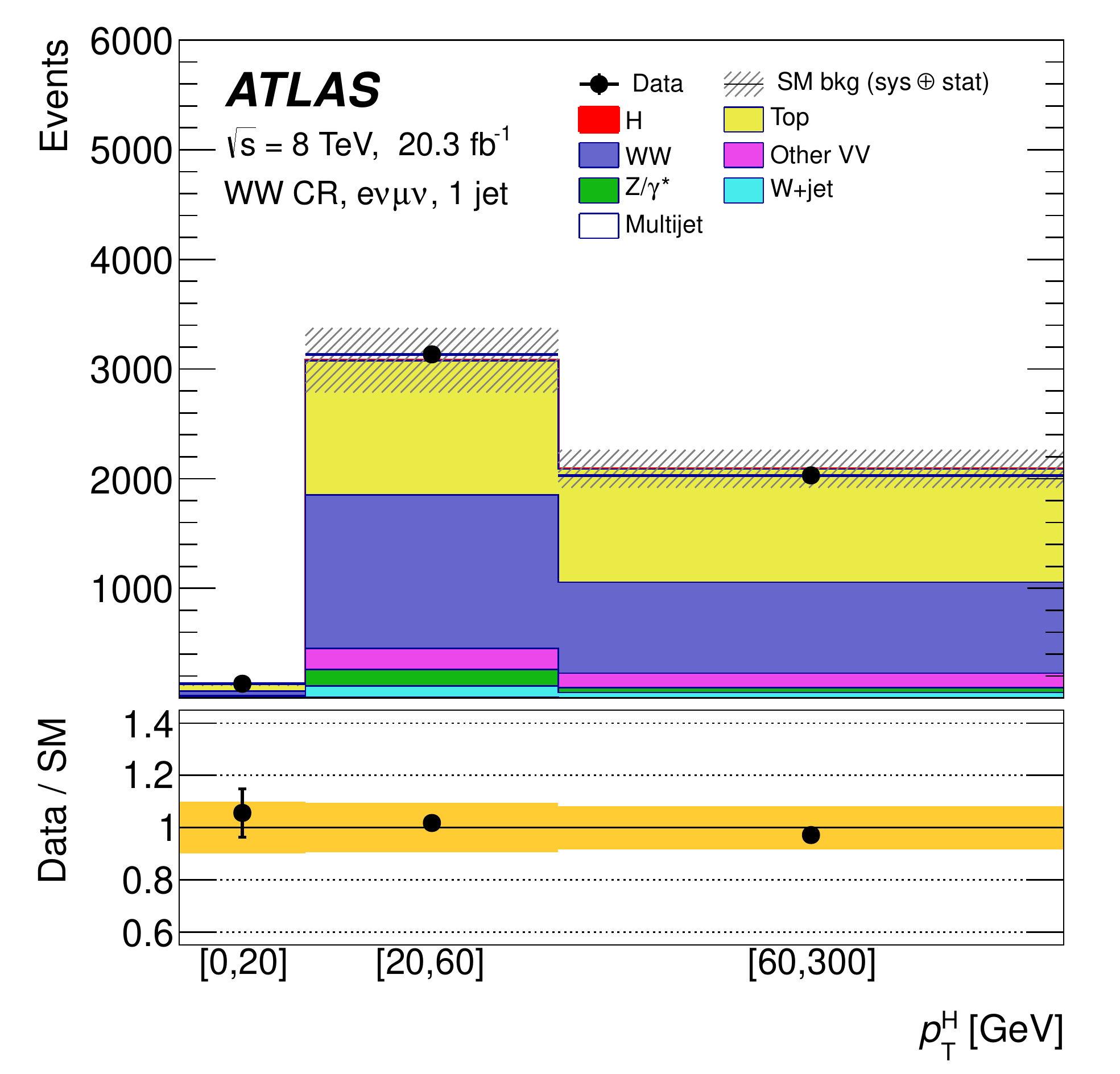}
  }
  \caption{ \label{fig:WWCR}  Observed distributions of (a) $\yll$ in the $\ZeroJet$ $WW$ CR and (b) $\pTH$ in the $\OneJet$ $WW$ CR,
    with signal and background expectations.  Relevant background normalisation factors have been applied.
		  The SM Higgs boson signal prediction shown is summed over all production processes.
      The hatched band in the upper panel and the shaded band in the lower panel show the sum in quadrature of statistical 
    and systematic uncertainties of the prediction.  
    }
\end{figure}

The top-quark background normalisation is estimated using control regions for all $\Njet$, and the shapes
of the distributions other than $\Njet$ are taken from MC simulation.  The $t\bar{t}$ and single-top (i.e. $Wt$) 
backgrounds are treated together and the normalisation factor determined from the CR yield is applied to their sum.  
In the $\ZeroJet$ category, the normalisation is derived from an 
inclusive sample of events meeting all of the lepton and $\met$ preselection criteria but with no requirements 
on the number of jets, in which the majority of events contain top quarks.  The efficiency of the $\ZeroJet$ signal 
region selection is modelled using MC simulation. To reduce the uncertainty on the efficiency of the jet veto, the fraction of $b$-tagged events which have no additional jets is measured in a data sample with at least one $b$-tagged jet and compared to the fraction predicted by simulation. The efficiency of the jet veto is corrected by the square of the ratio of the measured fraction over the predicted one to account for the presence of two jets in $t\bar{t}$ production.
In the $\OneJet$ category, the normalisation of the top-quark 
background is determined from a control region distinguished from the signal region by requiring that the 
jet is $b$-tagged.  To reduce the effect of $b$-tagging systematic uncertainties, the extrapolation factor from the CR to 
the SR is corrected using an effective $b$-jet tagging scale factor derived from a control region with two 
jets, at least one of which is $b$-tagged.  In the $\TwoJet$ category, the number of top-quark events is 
sufficiently large that a CR with a $b$-jet veto can be defined using $\mll\,>\,80\GeV$.
The $\pTj$ distribution in the $\OneJet$ top-quark CR and the $\pTH$ distribution in the $\TwoJet$ top-quark CR are 
shown in Figure~\ref{fig:TopCR}.
\begin{figure}[tbp!]
  \centering
  \subfloat[$\pTj$, $\OneJet$]{
    \includegraphics[width=0.46\textwidth]{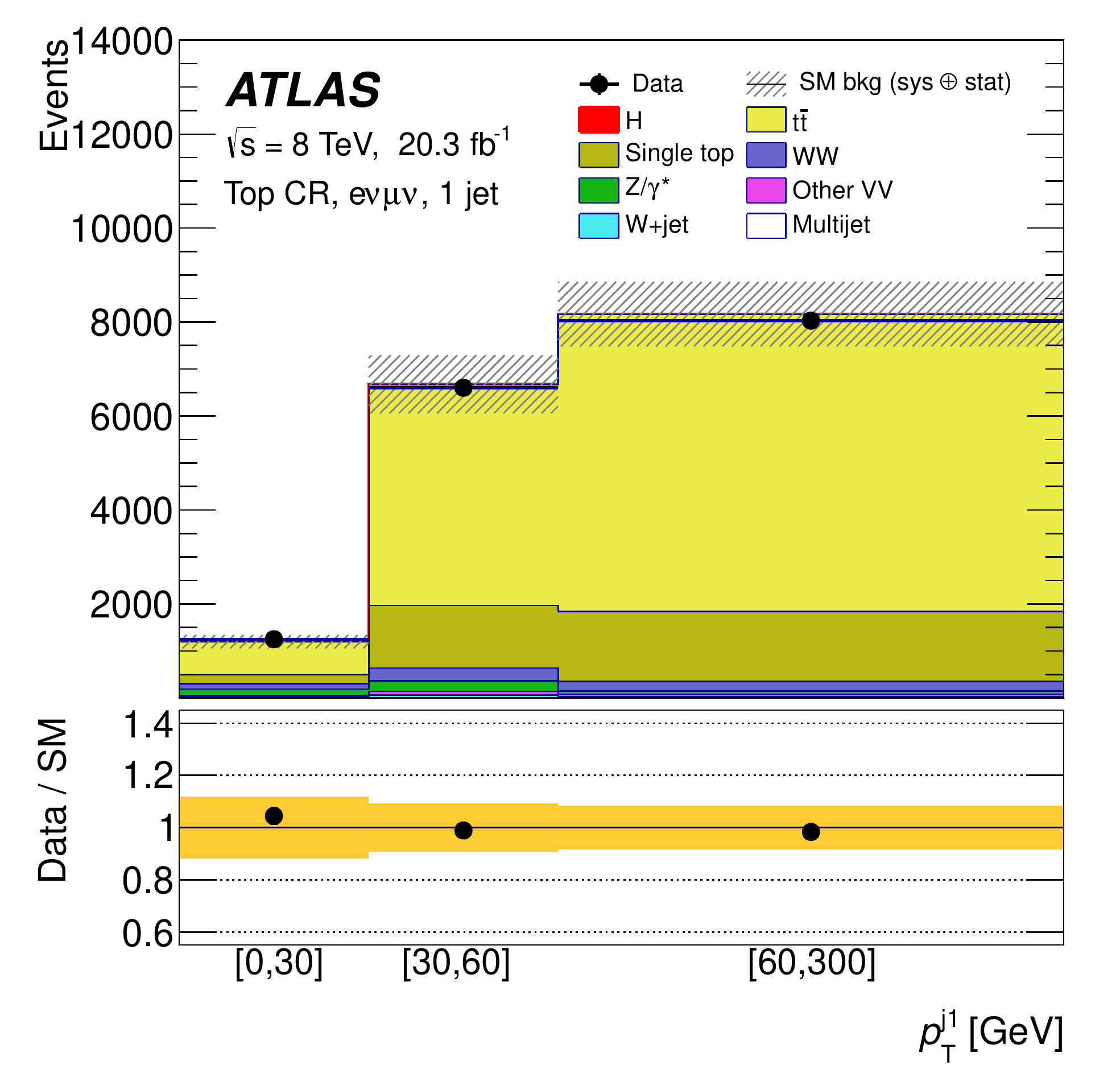}
  }
  \subfloat[$\pTH$, $\TwoJet$]{
	\includegraphics[width=0.46\textwidth]{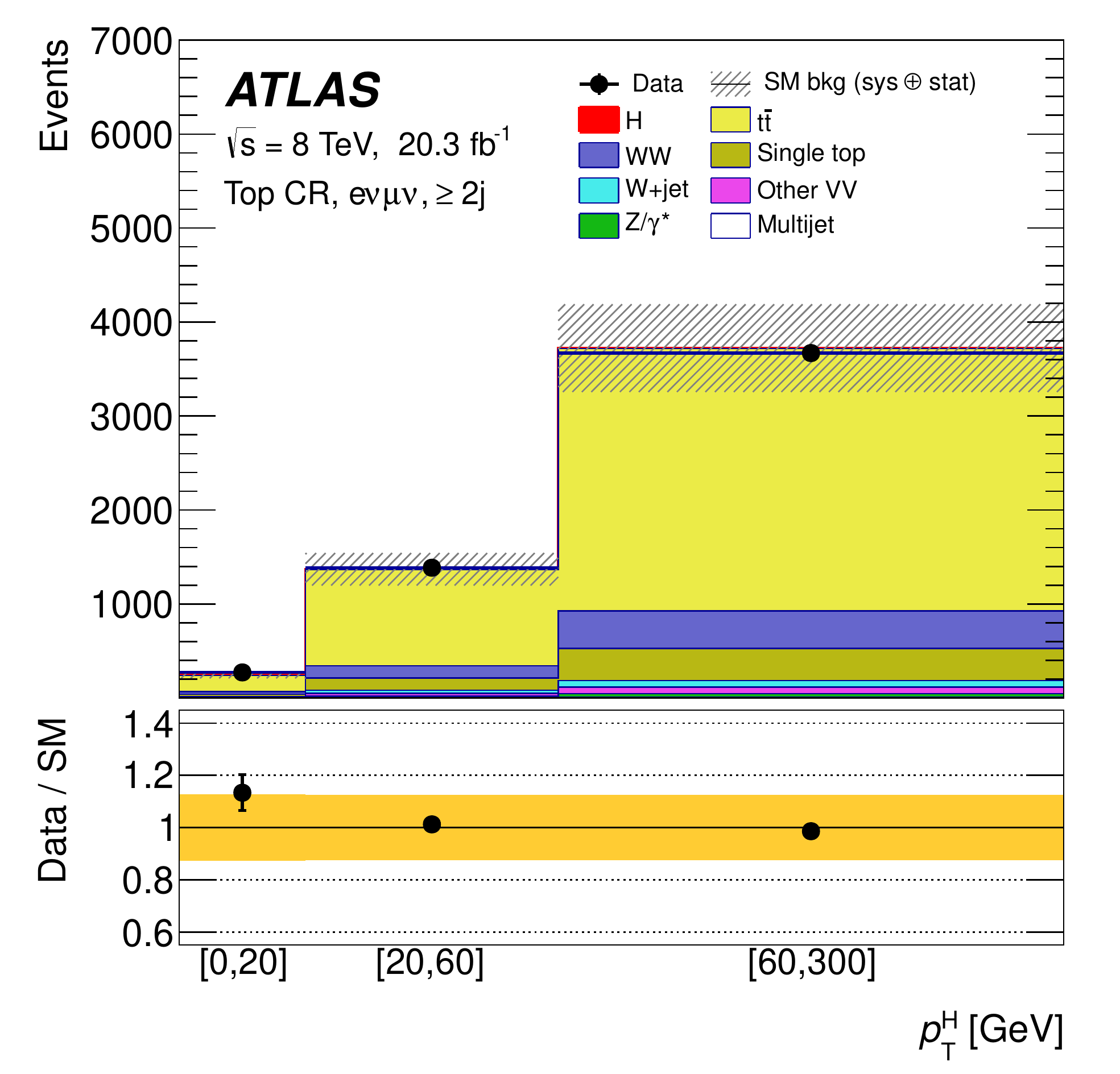}
  }
  \caption{ \label{fig:TopCR}  Observed distributions of (a) $\pTj$ in the $\OneJet$ top-quark CR and (b) $\pTH$ in the 
    $\TwoJet$ top-quark CR, with signal and background expectations.  Relevant background normalisation factors have been applied.
		  The SM Higgs boson signal prediction shown is summed over all production processes.
      The hatched band in the upper panel and the shaded band in the lower panel show the sum in quadrature of statistical 
    and systematic uncertainties of the prediction.  
    }
\end{figure}

The $\Wjets$ background contribution is estimated using a control sample of events in which one of the two lepton 
candidates satisfies the identification and isolation criteria used to define the signal sample (these lepton candidates 
are denoted ``fully identified''), and the other (``anti-identified'') lepton fails to meet the nominal selection criteria 
but satisfies a less restrictive one. Events in this sample are otherwise required to 
satisfy all of the signal-region selection criteria.  The $\Wjets$ contamination in the SR is determined by scaling the 
number of events in the control sample by an extrapolation factor measured in a $\Zjets$ data sample.
The extrapolation factor is the ratio of the number of fully identified leptons to the number of anti-identified 
leptons, measured in bins of anti-identified lepton $\pT$ and $\eta$.  To account for differences between the composition of jets 
associated with $W$- and $Z$-boson production, the extrapolation factors are measured in simulated $\Wjets$ and 
$\Zjets$ events. The ratio of the two extrapolation factors is applied as a multiplicative correction to the extrapolation factor 
measured in the $\Zjets$ data.  The background due to multijet events is determined similarly to the $\Wjets$ background, using a 
control sample that has two anti-identified lepton candidates, but otherwise satisfies the SR selection criteria. 
The extrapolation factor is constructed from data events dominated by QCD-produced jet activity, and is applied to both 
anti-identified leptons.

The background from diboson processes other than $WW$, primarily from $W\gamma^*$, $W\gamma$, and $WZ$ events,
is normalised in the $\ZeroJet$ and $\OneJet$ categories using a control region identical
to the signal region except that the leptons are required to have the same sign.  The number and properties
of same-sign and opposite-sign dilepton events produced by $W\gamma^{(*)}$ and $WZ$ are almost identical.  
In the $\TwoJet$ analysis, this same-sign sample is too small to be used as a control region, and
the background is estimated from the predicted inclusive cross sections and MC acceptance alone.
For all $\Njet$, the MC simulation is used to predict the shapes of the distributions to be unfolded.
Figure~\ref{fig:SSandZttCR}(a) shows the distribution of $\yll$ in the $\ZeroJet$ same-sign control region.

The $\Ztt$ background normalisation is derived from control regions, and the shape is derived from MC, for 
all three signal regions.  The small contributions from $\Zee$ and $\Zmm$, including $Z\gamma$, are estimated from
MC simulation and the predicted cross sections, as described in Section~\ref{sec:mcpred}.
Figure~\ref{fig:SSandZttCR}(b) shows the distribution of $\pTH$ in the $\Ztt$ control region with $\TwoJet$.
\begin{figure}[tbp!]
  \centering
  \subfloat[$\yll$, $\ZeroJet$ $VV$ CR]{
	\includegraphics[width=0.46\textwidth]{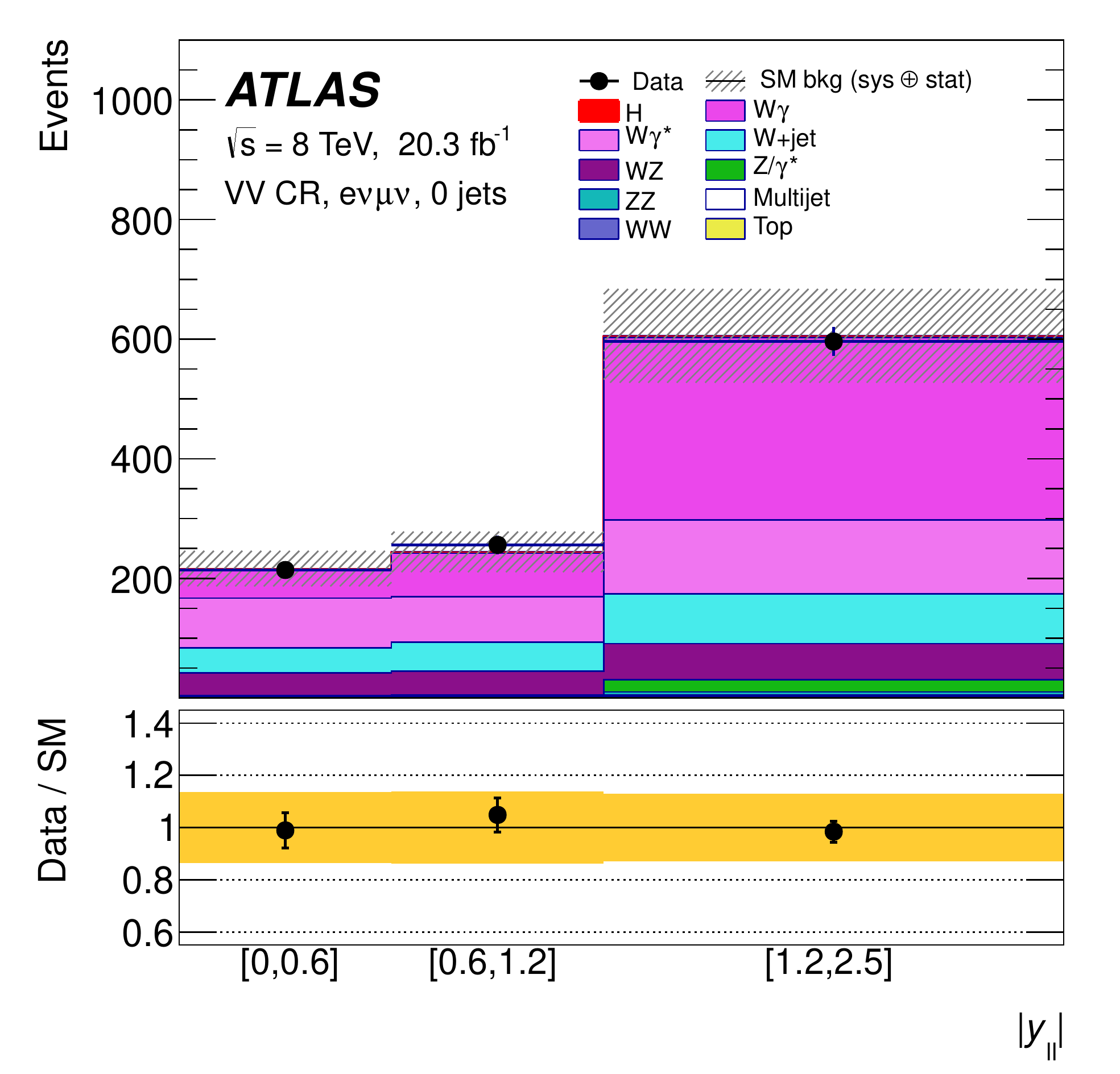}
	\label{subfig:yllVVCR}
  }
  \subfloat[$\pTH$, $\TwoJet$ $\Ztt$ CR]{
	\includegraphics[width=0.46\textwidth]{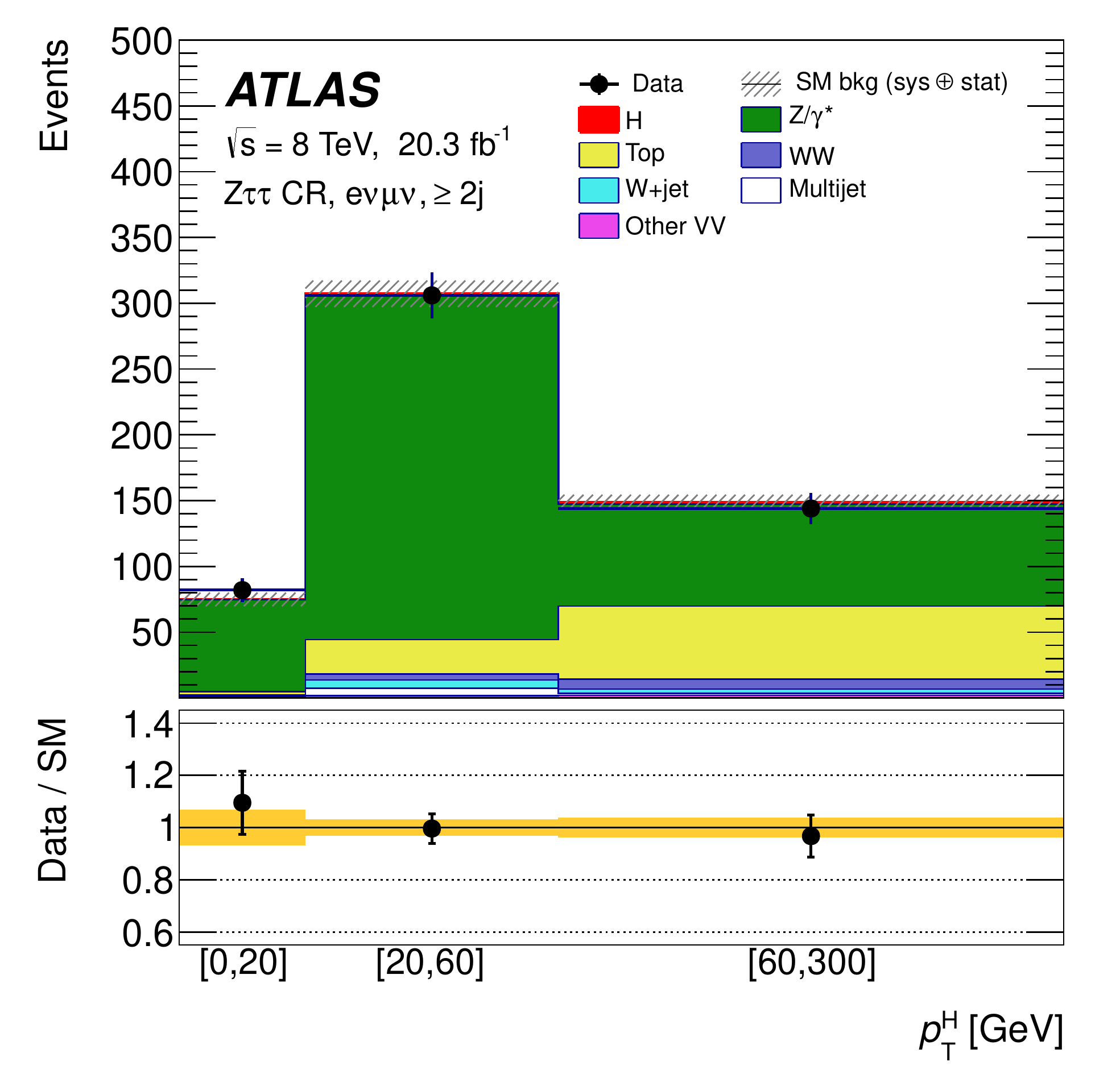}
	\label{subfig:pthZttCR}
  }
\caption{ \label{fig:SSandZttCR}  Observed distributions of (a) $\yll$ in the $\ZeroJet$ same-sign ($VV$) CR and (b)
   $\pTH$ in the $\TwoJet$ $\Ztt$ CR, with signal and background expectations.  Relevant background normalisation factors have been applied.
		  The SM Higgs boson signal prediction shown is summed over all production processes.
    The hatched band in the upper panel and the shaded band in the lower panel show the sum in quadrature of statistical 
    and systematic uncertainties of the prediction.  
}
\end{figure}

Each control region is designed for the calculation of a normalisation factor (NF) for a particular target process, 
The NF is defined as $(N - B')/B$, where $N$ is the number of data events observed 
in the control region, $B$ is the expected background yield in the CR for the target process based on the predicted cross section and 
acceptance from MC simulation, and $B'$ is the predicted 
yield from other processes in the control region.  The CRs have a small contribution from the signal process, which is 
normalised to the SM expectation.  The effect of this choice is negligible.
The normalisation of each background associated with a CR is scaled by the 
corresponding NF.  All NFs used are given in Table~\ref{tab:CR_NFs}, along with their statistical uncertainties.
These are included in the statistical uncertainties of the final results. 
The value of the \ZeroJet\ $WW$ NF has been studied in detail~\cite{HIGG-2013-13}; its deviation from unity is due to the modelling of the jet veto and higher-order corrections on the prediction of the $WW$ cross section.
A newer calculation of the inclusive $WW$ cross section, with NNLO precision in $\alpha_{\rm S}$~\cite{Gehrmann:2014fva}, moves
the NF closer to unity, compared to the one shown here, as described in Ref.~\cite{STDM-2013-07}.
\begin{table*}[tb!]
\centering
\caption{
Background normalisation factors (NFs) obtained from the control regions, for different background contributions and $\Njet$ 
  categories. The uncertainty quoted is the statistical uncertainty; systematic uncertainties on the predicted yield, not shown, restore
  compatibility of the NF with unity but do not directly enter the analysis because they are replaced by extrapolation uncertainties.
   A dash (`-') indicates that there is no control region corresponding to that background.
}
\begin{tabular}{lllll}
\dbline
Control Regions	& $WW$ & 	Top 	     & \Ztt\      & Other $VV$ \\
	  \sgline
\ZeroJet 	& 1.22 $\pm$ 0.03         & 1.08 $\pm$ 0.02         & 0.99 $\pm$ 0.02    	&  0.92 $\pm$ 0.07 \\
\OneJet 	& 1.05 $\pm$ 0.05        & 1.06 $\pm$ 0.02         & 1.06 $\pm$ 0.04        	&  0.96 $\pm$ 0.12 \\
\TwoJet 	& -         & 1.05 $\pm$ 0.03         & 1.00 $\pm$ 0.09       	& -  \\
  \dbline
\end{tabular}
\label{tab:CR_NFs}
\end{table*}

\section{Reconstructed yields and distributions}
\label{sec:yield}
The numbers of expected and observed events satisfying all of the signal region selection criteria are shown in 
Table~\ref{tab:eventyield}.  The numbers of expected signal and background events 
are also shown, with all data-driven corrections and normalisation factors applied.  
In each category, the background-subtracted number of events, corresponding to the observed yield of signal events, is significantly different from zero.
Taking into account the total statistical and systematic uncertainties, these yields are in agreement with those reported in 
Ref.~\cite{HIGG-2013-13} and with expectations from SM Higgs boson production through gluon fusion.

The four distributions under study: $\Njet$, $\pTH$ (reconstructed as $\pT(\ell\ell\met)$), $\yll$, and
$\pTj$ are shown in Figure~\ref{fig:measureddistributions}.  For presentation purposes, the reconstructed distributions 
are combined over the three signal regions, with the uncertainties combined accounting for correlations.  In the $\pTj$ 
distribution, $\ZeroJet$ events are all in the first bin, $\pTj\,<\,30\GeV$, by construction because of the definition of the jet counting.  
The composition of the background is shown, to illustrate how it varies as a function of the quantities being measured.  The $WW$ background decreases as a function of the number of jets,
and the top-quark background increases, as can also be seen in 
Table~\ref{tab:eventyield}. For the $\pTH$ and $\pTj$ distributions, the $WW$ background decreases with $\pT$ while the top-quark background increases. 
The background composition does not vary substantially as a function of $\yll$.  

\begin{table}[t]
\centering
\caption{Predicted and observed event yields in the three signal regions. Predicted numbers are given with their statistical 
  (first) and systematic (second) uncertainties evaluated as described in Section~\ref{sec:uncerts}.  
    The ``Non-ggF $H$'' row includes the contributions from VBF and $VH$ with $H{\rightarrow\,}WW^{\ast}$ and from $H\to\tau\tau$.
    The total background in the third-from-last row is the sum of these and of all other backgrounds.  }
\scalebox{0.95}{
\begin{tabular}{lr@{$\,\pm\,$}r@{$\,\pm\,$}rr@{$\,\pm\,$}r@{$\,\pm\,$}rr@{$\,\pm\,$}r@{$\,\pm\,$}r}
\dbline
 & \multicolumn{3}{c}{\ZeroJet} & \multicolumn{3}{c}{\OneJet} & \multicolumn{3}{c}{\TwoJet} \\
   \sgline
 Non-ggF $H$ 		&   $2.2$ &  $0.2$ &  $0.2$ 	&   $7.1$ & $0.3$ & $ 0.5$ 	&   $8.2$ & $0.3$ &  $0.4$ \\
 $WW$ 			& $686$ & $19$ & $43$	& $153$ & $7$ & $13$	&  $44$ & $1$ & $11$ \\
 Other $VV$ 		&  $88$ &  $3$ & $12$	&  $44$ & $3$ & $11$	&  $21.6$ & $1.6$ &  $3.3$ \\
Top 			&  $60.2$ &  $1.5$ &  $3.8$	& $111.2$ & $2.7$ & $ 8.2$ 	& $164$ & $2$ & $16$ \\

 $Z/\gamma^*$ 		&   $8.7$ &  $2.3$ &  $2.3$	&   $6.2$ & $1.3$ & $ 2.2$	&   $7.3$ & $1.5$ &  $2.2$ \\ 
 $W$+jets 		&  $90$ &  $2$ & $21$	&  $33.5$ & $2.0$ & $ 7.6$	&  $16.9$ & $1.2$ &  $3.9$ \\
 Multijet 			&   $1.3$ &  $0.5$ &  $0.5$	&   $0.7$ & $0.2$ & $ 0.3$	&   $0.9$ & $0.1$ &  $0.4$ \\
 \sgline
 Total background	& $936$ & $21$ & $41$ 		& $355  $ & $9  $ & $12$ 	&   $263$ & $6$ & $9$ \\
 Observed               & \multicolumn{3}{c}{$1107$} 
 			& \multicolumn{3}{c}{$414$} 
 			& \multicolumn{3}{c}{$301$}  \\
 \sgline
 Observed $-$ background	& $171$ & $39$ & $41$		& $59$ & $22$ & $12$		& $38$ & $18$ & $9$ \\ 
 \sgline	
  ggF $H$ 		& $125.9$ & $0.4$ & $5.7$	&  $43.4$ & $0.2$ & $ 1.7$ 	&  $17.6$ & $0.2$ &  $1.4$ \\	
 \dbline
\end{tabular}}
\label{tab:eventyield}
\end{table}

\begin{figure}[!tbp]
  \centering
  \subfloat[$\Njet$]{
    \includegraphics[width=0.46\textwidth]{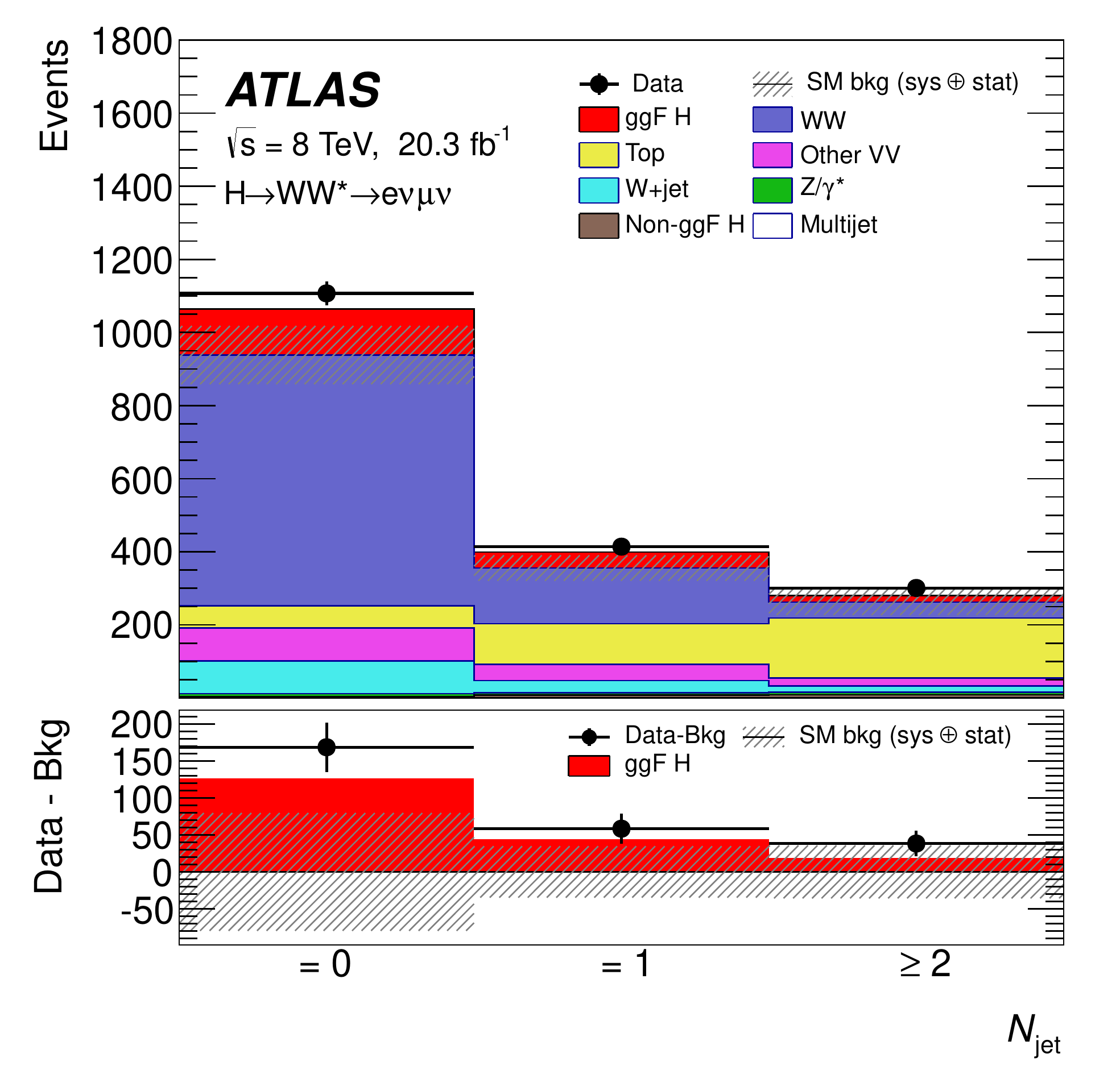}

  }
  \subfloat[$\pTH$]{
    \includegraphics[width=0.46\textwidth]{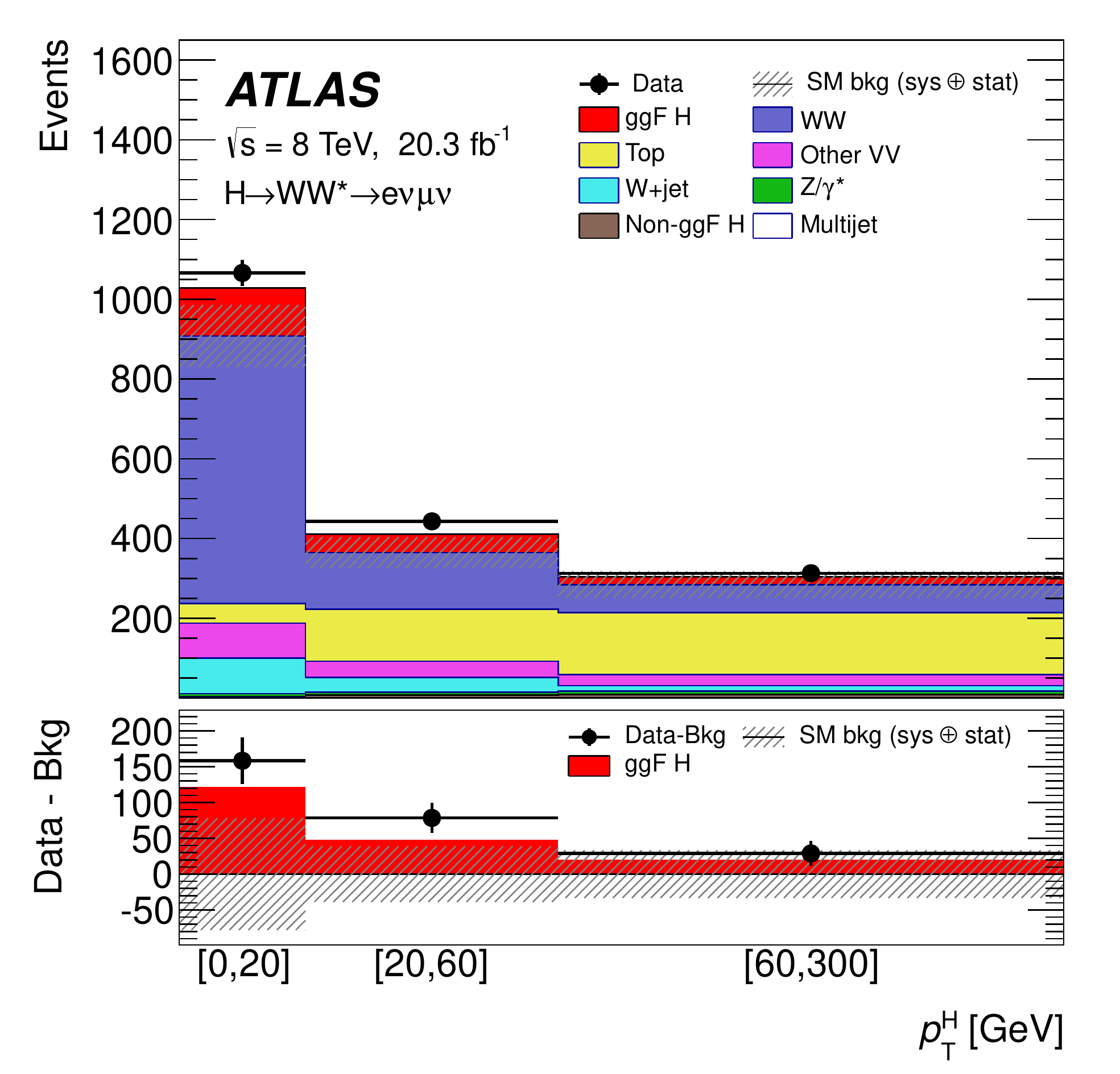}
  }
  \newline
  \subfloat[$\yll$]{
    \includegraphics[width=0.46\textwidth]{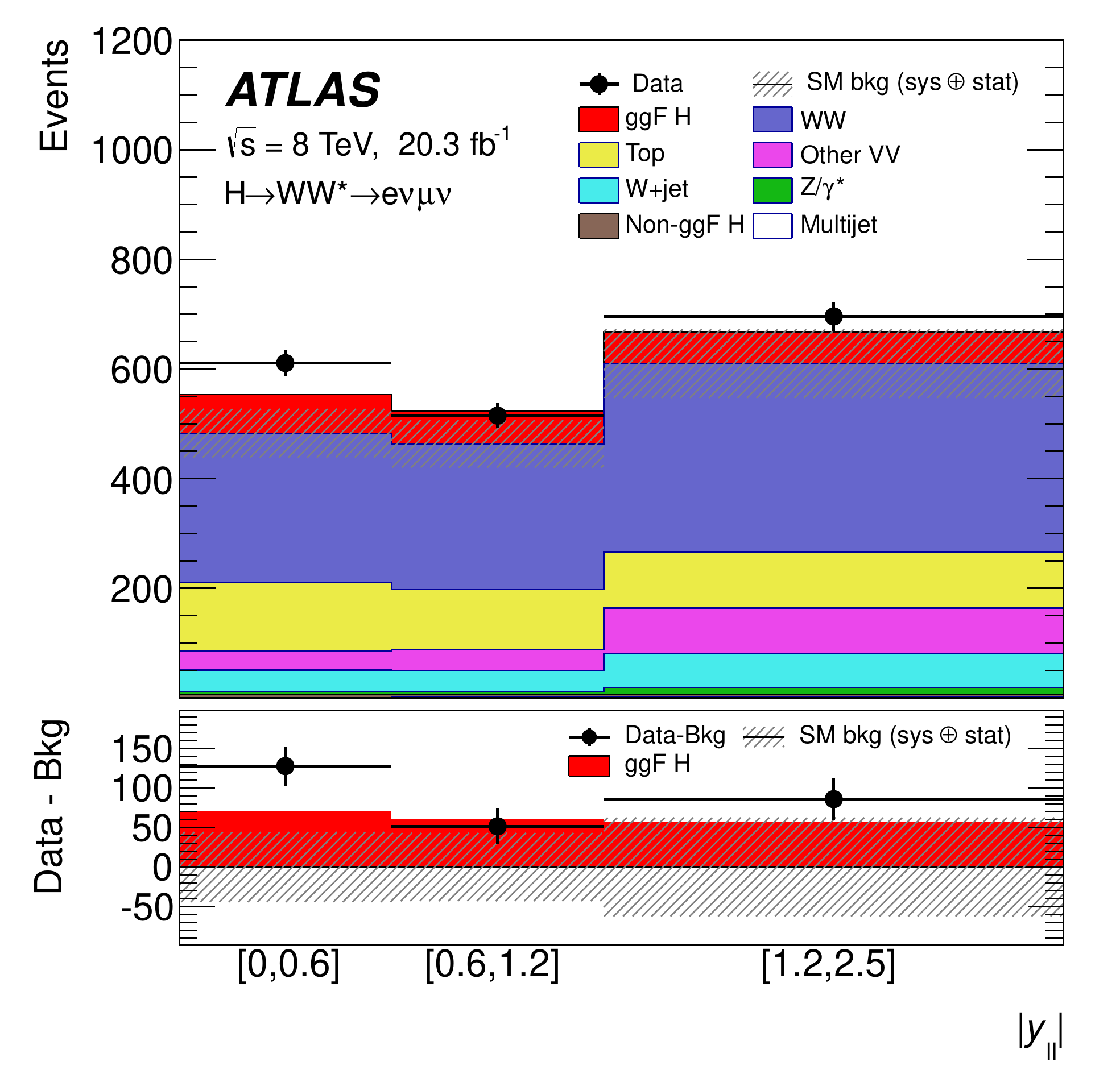}
  }
  \subfloat[$\pTj$]{
    \includegraphics[width=0.46\textwidth]{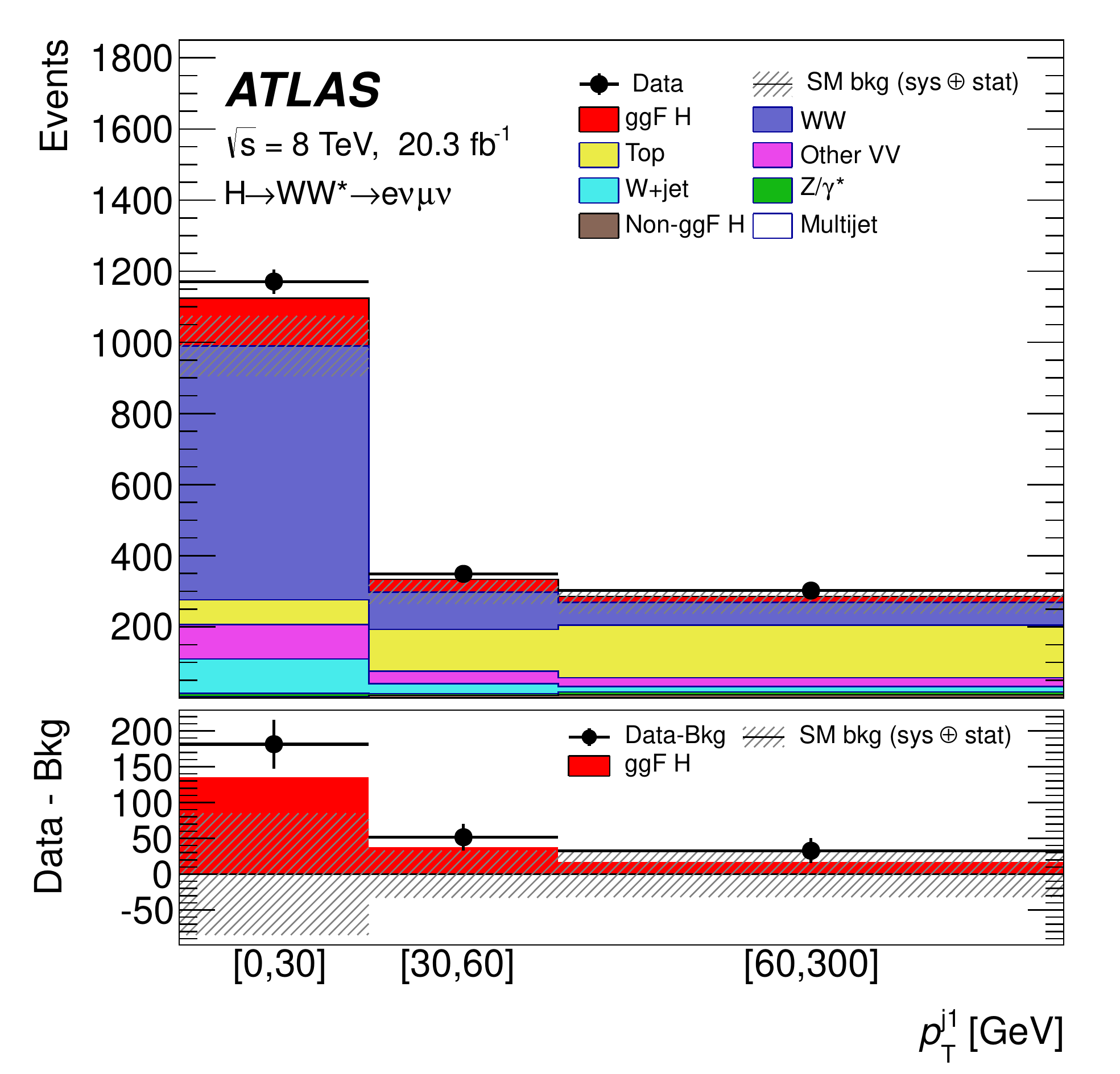}
  }
  \caption{Observed distributions of (a) $\Njet$, (b) $\pTH$, (c) $\yll$, and (d) $\pTj$ with signal and background expectations, combined over the \AllJet\ signal-region categories.   
		The background processes are normalised as described in Section~\ref{sec:background}.  
		  The SM Higgs boson signal prediction shown is summed over all production processes.
In the $\pTj$ distribution, $\ZeroJet$ events are all in the first bin by construction because of the definition of the jet thresholds used
to define the signal regions.
  The hatched band shows the sum in quadrature of statistical and systematic uncertainties of the sum of the backgrounds. 
  \label{fig:measureddistributions}
    }
\end{figure}

\section{Fiducial region and correction for detector effects}
\label{sec:unfolding}
Each of the reconstructed distributions is corrected for detector effects and resolution to extract the differential cross sections for the ggF Higgs boson signal.
All differential cross sections are shown in a fiducial region defined based on objects at particle level, to reduce the model dependence of the results.
The particle objects and the definition of the fiducial region are described in Section~\ref{sec:fidvol}. In Section~\ref{sec:corr}, 
the correction procedure is discussed. 

\subsection{Definition of the fiducial region}
\label{sec:fidvol}

The fiducial selection is designed to replicate the analysis selection described in Section~\ref{sec:selection} as closely as possible at particle level, before the simulation of detector effects.
In this analysis, measurements are performed in three signal-region categories differing in the number of jets in the event. In order to present results with events from all
categories, the fiducial selection only applies a selection common to all categories and using the leptons and missing transverse momentum in the final state. The criteria are summarised in Table~\ref{tab:fidsel}.

The fiducial selection is applied to each particle-level lepton, defined as a final-state electron or muon. Here, electrons or muons from hadron decays and $\tau$ decays are rejected.
The lepton momenta are corrected by adding the momenta of photons, not originating from hadron decays, within a cone of size $\Delta R = 0.1$ around each lepton; these photons arise predominantly from final-state-radiation. Selected leptons are required to satisfy the same kinematic
requirements as reconstructed leptons. 
A selected event has exactly two different-flavour leptons with opposite charge.

The missing transverse momentum $\vmet$ is defined as the vector sum of all final-state neutrinos excluding those produced in the decays of hadrons and $\tau$'s. 

Particle-level jets are reconstructed using the anti-$k_{t}$ algorithm, implemented in the \textsc{FastJet}\xspace package~\cite{Cacciari:2011ma}, with a radius parameter of $R = 0.4$. For the clustering, all stable particles with a mean lifetime greater than 30~ps are used, except for electrons, photons, muons, and neutrinos not originating from hadron decays. Selected jets are required to have $\pT\,{>}\,25\GeV$ if $|\eta| < 2.4$ or $\pT\,{>}\,30\GeV$ if $2.4 \leq |\eta| < 4.5$. 

Selected events pass all preselection requirements introduced in Section~\ref{sec:selection} and the $\hwwenmn$ topology selection on $\Delta\phi_{\ell\ell}$ and $m_{\ell\ell}$. 
The $\mT$ thresholds are not applied in the fiducial region since the shape of the $\mT$ distribution at reconstruction level differs significantly from the shape of the distribution at particle level. 
All selection requirements applied are summarised in Table~\ref{tab:fidsel}. 
For a SM Higgs boson the acceptance of the fiducial region with respect to the full phase space of $\hwwenmn$ is 11.3\%. 

\begin{table}[t!]
  \centering
  \caption{
   Summary of the selection defining the fiducial region for the cross-section measurements. The momenta of the electrons and muons 
     are corrected for radiative energy losses by adding the momenta of nearby photons, as described in the text.  
  }
  \begin{tabular}{ll}
  \dbline
  \multicolumn{2}{c}{\bf Object selection} \\
  Electrons   & $\pT\,{>}\,15\GeV$, $|\eta| < 1.37$ or $1.52 < |\eta| < 2.47$ \\
  Muons 	& $\pT\,{>}\,15\GeV$, $|\eta| < 2.5$ \\
  Jets		& $\pT\,{>}\,25\GeV$ if $|\eta| < 2.4$, $\pT\,{>}\,30\GeV$ if $2.4 \leq |\eta| < 4.5$ \\
  \sgline
  \multicolumn{2}{c}{\bf Event selection} \\
  \multirow{3}{*}{Preselection} & $\pTlead(\ell) > 22 \GeV$ \\
  			& $m_{\ell\ell} > 10 \GeV$ \\
			& $\MET > 20 \GeV$ \\
  \multirow{2}{*}{Topology} & $\Delta\phi_{\ell\ell}<1.8$ \\
  			& $m_{\ell\ell}<55 \GeV$ \\
  \dbline
  \end{tabular}
  \label{tab:fidsel}
\end{table} 

\subsection{Correction for detector effects}
\label{sec:corr}

To extract the differential cross sections, the measured distributions, shown in Figure~\ref{fig:measureddistributions}, are corrected for detector effects and extrapolated to the fiducial region. 
For the corrections, the reconstructed distributions of the different jet-binned signal-region categories are not combined, but instead are simultaneously corrected for detector effects as a function of the variable under study and the number of jets.
Thus, the correlation of the variable under study with $\Njet$ is correctly taken into account.
Final results are presented integrated over all values of $\Njet$ for the $\pTH$, $|\yll|$ and $\pTj$ variables. 

In the following, each bin of the reconstructed distribution is referred to by the index $j$, while each bin of the 
particle-level distribution is referred to by the index $i$. The correction itself is done as follows:
\begin{linenomath}
\begin{equation}
 N_{i}^{\mathrm{part}} = \frac{1}{\varepsilon_{i}}  \cdot \sum_j \left ( M^{-1} \right)_{ij} \cdot f_{j}^{\textrm{reco-only}} \cdot (N_{j}^{\mathrm{reco}} - N_{j}^{\mathrm{bkg}}) , 
\end{equation}
\end{linenomath}
where $N_{i}^{\mathrm{part}}$ is the number of particle-level events in a given bin $i$ of the particle-level distribution in the fiducial 
region.  The quantity $N_{j}^{\mathrm{reco}}$ is the number of reconstructed events in a given bin $j$ of the reconstructed distribution in 
the signal region, and $N_{j}^{\mathrm{bkg}}$ is the number of background events in bin $j$ estimated as explained in 
Section~\ref{sec:background}. The correction factor $f_{j}^{\textrm{reco-only}}$, the selection efficiency $\varepsilon_{i}$, and 
the migration matrix $M_{ij}$ are discussed below. To evaluate the cross section in particle-level bin $i$, it is also 
necessary to take the integrated luminosity and the bin width into account.

The migration matrix accounts for the detector resolution and is defined as the probability to observe an event in bin $j$ when its particle-level value is located in bin $i$. The migration matrix is
built by relating the variables at reconstruction and particle level in simulated ggF signal events that meet both the signal-region and fiducial-region selection criteria. To properly account for the migration of events between the different signal-region categories, the migration matrix accounts for the migrations within one distribution, as well as migrations between different values of $\Njet$. 
The inverse of the migration matrix is determined using an iterative Bayesian unfolding procedure~\cite{DAgostini:1994zf} with two iterations. 

The selection efficiency $\varepsilon_{i}$ is defined as an overall efficiency, combining reconstruction, identification, isolation, trigger and selection, including also the differences between the fiducial and the signal region selection. It is derived from MC simulation and its values are in the range 0.14 to 0.43 for all variables. 
Events in the fiducial region that are not selected in the signal region are taken into account by $\varepsilon_{i}$.

Events outside the fiducial region may be selected in a signal region owing to migrations.
Such migrations are accounted for via the correction factor $f_{j}^{\textrm{reco-only}}$, which is derived from MC simulation. Reconstructed $\hww$ events where the $W$ boson decays into $\tau\nu$ and the $\tau$ lepton decays leptonically are not included in the fiducial region, but are accounted for also with the same procedure. The correction factor $f_{j}^{\textrm{reco-only}}$ is in the range 0.84 to 0.92 for all variables.

\section{Statistical and systematic uncertainties}
\label{sec:uncerts}
Sources of uncertainty in the differential cross sections can be grouped into
five categories: statistical uncertainties, experimental systematic uncertainties, theoretical systematic uncertainties in the signal model, uncertainties arising from the correction
procedure, and theoretical systematic uncertainties in the background model. These uncertainties affect the analysis through the background 
normalisation, the background shape, the migration matrix, the selection efficiency, and the correction factor.  

The effect of each systematic uncertainty is analysed by repeating the full analysis for the variation
in the signal, background, or experimental parameter.  For experimental uncertainties, the migration matrix, 
selection efficiency, correction factor, and background estimation are varied simultaneously.  For uncertainties 
that only apply to the background processes, the nominal migration matrix, selection efficiency, and correction factor are used. 
The total uncertainty in the result from any individual source of uncertainty
is taken as the difference between the shifted and the nominal result after the correction of detector effects.

The input uncertainties are summarised in this section. Their effect on the measured results, individually 
and collectively, are given with the results in the tables in Section~\ref{sec:results}.
The total uncertainty in each measurement bin is defined as the sum in quadrature of all uncertainty components.

\subsection{Statistical uncertainties}

The statistical uncertainties in the differential cross sections are estimated using pseudo-experiments. The 
content of each bin in the measured distribution is fluctuated according to a Poisson distribution. In each pseudo-experiment 
the background is subtracted and the correction for detector effects is performed. Then, the root mean square of the spread 
of the result in each bin is taken as the estimator of the statistical uncertainty. Values for the data statistical uncertainty are evaluated
using pseudo-experiments; the data statistical uncertainties in the presented measurement range from $17\%$ to $61\%$.

The uncertainty due to the statistics of the background MC samples 
is evaluated by fluctuating the bin contents of the background template using a Gaussian distribution 
with a width corresponding to the uncertainty in that bin. In case of the signal MC sample, the bins of the migration 
matrix, the selection efficiency, and the correction factor are fluctuated simultaneously. 
In each pseudo-experiment the correction for detector effects is performed using the respective fluctuated template. The 
root mean square of the spread of results of the pseudo-experiments is taken as the estimator of the uncertainty.

For results integrated over all values of $\Njet$, and for normalised results, each pseudo-experiment is integrated or normalised and the uncertainty is re-evaluated 
for the integrated (normalised) bin to take into account all correlations arising from bin migration.

The statistical uncertainties in the background normalisations from the data yields in the control regions
are calculated as the square root of the number of events observed. 

\subsection{Experimental systematic uncertainties}

Experimental systematic uncertainties arise primarily from object calibrations, such as the jet energy 
scale, and affect the subtracted background normalisation and shape as well as the migration matrix, the selection efficiency, and the correction factor.  
The variations used for the experimental uncertainties are identical to those 
of Ref.~\cite{HIGG-2013-13} and are not described here. The effect of these variations have been reevaluated in the context of this analysis. The dominant experimental uncertainties are those associated with the jet 
energy scale (JES) and resolution (JER), the lepton identification efficiencies, and the 
uncertainty in the extrapolation factor used to estimate the $\Wjets$ background. 
For each uncertainty, the upward and downward
variations are performed separately.  Each variation is applied simultaneously to the migration matrix, the 
selection efficiency, the correction factor, and the background subtraction so that correlations are correctly preserved. 
The background-subtracted yields are allowed to assume negative values under these variations.

\subsection{Systematic uncertainties in the signal model}
\label{sec:sigunc}

Theoretical uncertainties in the ggF signal model can affect the migration matrix, the selection efficiency, and the correction factor.
Sources of theoretical uncertainty in the signal acceptance are the choice of QCD renormalisation and factorisation
scales, PDF, parton shower/underlying event (PS/UE) model, and matrix-element generator.
It was shown in Ref.~\cite{HIGG-2013-13}
that the theoretical uncertainty in the signal acceptance is dominated by the PS/UE model. This uncertainty is evaluated by 
constructing the migration matrix and the correction factors with \POWHEG+\HERWIG and \POWHEG+\PythiaEight and applying both sets in the detector correction. The results with each of the simulations
are then compared for each of the measured distributions. The full difference between the distributions is taken as an uncertainty, which is at the level of a few percent. 

In addition to the uncertainty in the signal acceptance, an uncertainty in the theoretical predictions of the exclusive ggF $H+n$-jet cross sections is assigned. 
The uncertainties in the exclusive cross sections are evaluated using the jet-veto efficiency method~\cite{Dittmaier:2012vm,Andersen:2014efa}. 
Here, uncertainties due to renormalisation, factorisation, and resummation scale choices in the analytical calculations are taken into account. The correlations of the uncertainties in the different
$H+n$-jet cross sections are determined using a covariance matrix as described in Ref.~\cite{Boughezal:2013oha}. 
To evaluate the effect this uncertainty has on the migration matrix, the selection
efficiency, and the correction factor, the particle-level $\Njet$ distribution in the signal ggF MC sample 
is reweighted to account for the uncertainties in the exclusive $H+n$-jet cross sections 
  and the correlations between them. Then, the reconstructed distribution of the reweighted ggF signal MC sample is unfolded for each variable to evaluate the change arising from the uncertainty in the exclusive ggF $H+n$-jet cross sections. The contribution of this uncertainty to the differential distributions is a few percent for $\pTH$ and $\pTj$ and negligible for $\yll$.

\subsection{Systematic uncertainty in the correction procedure}
\label{sec:unfunc}

The ggF signal simulation is used to build the migration matrix and can bias the result of the correction procedure. This bias is partly evaluated with the uncertainties in the SM prediction of the signal determined in Section~\ref{sec:sigunc}.
To evaluate this bias independently of the SM prediction and its uncertainty, the simulated ggF signal sample is reweighted to reproduce the amount of disagreement in shape between the reconstructed simulated distribution and the background-subtracted measured
distribution. For this reweighting, only the nominal distributions are compared; uncertainties are not taken into account.
The reweighted reconstructed distribution is then corrected for detector effects using the nominal migration matrix. The difference between the 
corrected distribution and the reweighted simulated particle-level distribution is taken as an uncertainty in the correction procedure. 
The resulting uncertainty is smaller than $5\%$ in each measurement bin.

\subsection{Systematic uncertainties in the background model}
\label{sec:bgsys}

Systematic uncertainties in the background model are evaluated by comparing the background predictions as
evaluated under different conditions.  For the dominant $WW$ and top-quark backgrounds, shape uncertainties 
in each measured distribution are considered in addition to normalisation uncertainties.
For the backgrounds normalised by a control region, the normalisation uncertainty is derived by varying
the extrapolation factor, and for backgrounds estimated directly from the MC simulation, such as 
the $WW$ background in the $\TwoJet$ signal region, the systematic uncertainty is derived by varying the 
full event yield in the SR rather than an extrapolation factor, and accounts for the uncertainty in the cross 
section and acceptance.

The nominal MC sample used to model the $WW$ background yield for the \ZeroOneJet categories is 
\POWHEG+\PythiaSix.  The theoretical uncertainties assessed are:
\begin{itemize}
\item QCD scales, by independently varying the values of the renormalisation
scale $\mu_{\rm R}$ and the factorisation scales $\mu_{\rm F}$, in \aMCATNLO\ calculations~\cite{mg5aMCatNLO}.  Both scales are independently
multiplied by a factor of 2.0 or 0.5 relative to the nominal value $\mu_{0} = m_{WW}$, where $m_{WW}$ is the invariant mass 
of the $WW$ system, while maintaining the constraint $0.5\,\le\,\mu_{\rm R}/\mu_{\rm F}\,\le\,2$.
\item PDF uncertainties, from the envelope of the CT10 68\% CL eigenvectors added in quadrature with the 
maximal difference between the results obtained with CT10 and those obtained with either MSTW~\cite{Martin:2009iq} or NNPDF~\cite{Ball:2012cx}.
\item The choice of parton-shower and underlying-event models (PS/UE), by comparing the
nominal {\POWHEG} prediction interfaced with \PythiaSix and \HERWIG.
\item The choice of matrix-element generator, by comparing the nominal \POWHEG\ to \aMCATNLO, both
interfaced with \HERWIG.
\end{itemize}
The normalisation uncertainties are summarised in Table~\ref{tab:ww_theory_uncerts}.  
These are all varied in a correlated way for the $\ZeroJet$ and $\OneJet$ signal regions.  
Each source is also considered as a shape uncertainty, except for the PDF uncertainty, which is
much smaller than the others.  The changes observed are typically 1--10\% for $\pTH$ and
$\pTj$, and less than $1\%$ for $\yll$.  The largest changes observed are from the effect the PS/UE variation 
has on $\pTH$ and occur in sparsely populated bins, 50\% for $\ZeroJet$ events with $\pTH > 60\GeV$ and 30\% for
$\OneJet$ events with $\pTH < 20\GeV$.  The shape and normalisation are varied simultaneously for 
the PS/UE and matrix-element-generator uncertainties.  
The QCD-scale uncertainties are taken from the variation exhibiting the largest difference from nominal,
which is $\mu_{\rm R}/\mu_0\,=\,2.0$ and $\mu_{\rm F}/\mu_0\,=\,2.0$ for both the $\ZeroJet$ and $\OneJet$ normalisation
uncertainties.  The shape uncertainties are set similarly, but the variation with the largest
difference to the nominal is not always the one driving the normalisation uncertainty.
The resulting shape uncertainties are not correlated with the normalisation uncertainties.

The theoretical uncertainties in the $WW$ background yield for the $\TwoJet$ category are evaluated 
similarly.  The QCD-scale uncertainty is evaluated by varying the renormalisation and factorisation scale $\mu$, which has the 
nominal value of $\mu_0 = m_{WW}$, in the range $0.5\,\le\,\mu/\mu_0\,\le\,2$ in \MADGRAPH~\cite{Alwall:2011uj}, and applying the relative uncertainty 
to the nominal \SHERPA prediction.  The choices of matrix-element generator and parton shower are varied 
together by comparing \MADGRAPH+\PythiaSix to \SHERPA.  Uncertainties in the predicted shape are 
also accounted for, and are between 1\% and 15\%.  
The larger uncertainties in the $\TwoJet$ category are due to the use of a different MC generator (multi-leg LO in QCD) and the absence of a CR.
For the same reasons, they are not correlated with the uncertainties in the $\ZeroJet$ and $\OneJet$ categories. 

\begin{table}[tbp]
\centering
\caption{Theoretical uncertainties (in \%) in the $WW$ background normalisation estimate in each signal region.
  	The relative sign between entries in a row indicates correlation or anti-correlation
	  among the $\ZeroJet$ and $\OneJet$ signal regions, as a single variation is applied simultaneously
	  to both of them.  The $\TwoJet$ uncertainties are treated as uncorrelated.
\label{tab:ww_theory_uncerts}}
\begin{tabular}{lrrr}
\dbline
  & $\ZeroJet$ & $\OneJet$ & $\TwoJet$ \\
\sgline
 QCD scales	& $-1.1$	& $-1.7$ & $+22$\phantom{.0} \\
\hline
 PDF 	&  $+0.6$	& $+0.6$	&  $+9.7$ \\
\hline
 PS/UE	& $-1.3$	& $-4.5$ & -- \\
\hline
 Generator 	& $+5.2$	& $+1.5$ & $+2.7$	\\
   \dbline
\end{tabular}
\end{table}

Shape and normalisation uncertainties in the top-quark background yield are evaluated following the procedure applied 
for the $WW$ background.  
The normalisation uncertainties for each signal region are summarised in Table~\ref{tab:top_theory_uncerts}.
In contrast to the $WW$ background, there is a non-negligible PDF shape uncertainty, which is evaluated by comparing CT10, MSTW, and NNPDF.  
For most uncertainty sources, the 
changes observed due to shape variations of the top-quark background are typically 5\% or smaller.  Exceptions are the PS/UE uncertainty for \ZeroJet events with 
$\pTH > 60\GeV$, which is about 12\%, and the PDF uncertainty in the \yll shape, which is up to 8\%.  
\begin{table}[tbp]
\centering
\caption{Theoretical uncertainties (in \%) in the top-quark background estimate in each signal region.
  	The relative sign between entries in a row indicates correlation or anti-correlation
	  among the signal regions.
\label{tab:top_theory_uncerts}}
\begin{tabular}{lrrr}
\dbline
  & \ZeroJet & \OneJet & \TwoJet \\
\sgline
 QCD scales     & $-1.2 $ & $-0.6 $ & $-0.8 $ \\
\hline
 PDF       & $+0.4 $ & $+2.2 $ & $+1.0 $ \\
\hline
 PS/UE      & $-0.6 $ & $+2.7 $ & $+4.5 $ \\
\hline
 Generator & $-4.1 $ & $-3.5 $ & $-1.1 $ \\
   \dbline
\end{tabular}
\end{table}

Very few MC-simulated events from the $\Ztt$ background pass the full SR and $\Ztt$ CR event selection, so the corresponding theoretical 
uncertainties are calculated with modified and reduced 
SR and CR selections, in order for the relevant comparisons to be made with sufficient 
statistical precision.  No shape uncertainty is assessed for the same reason, and the effect of any such uncertainty 
would be negligible due to the small contribution from this background. 
The $\pTZ$ distribution for $\ZeroJet$ events is reweighted using the ratio of data to MC simulation for
$\Zmm$ events produced with the same MC generator and PS/UE model, and the uncertainty in the reweighting procedure
is also included in the analysis.  The extrapolation uncertainty to the $WW$ control region is also evaluated, 
because the contribution of $\Ztt$ to that CR is not negligible.
As with the other backgrounds, each variation is applied simultaneously across all signal and control regions.

The systematic uncertainties in the contributions from $WZ$, $W\gamma$, $W\gamma^{\ast}$, and other small sources
of background are unmodified from Ref.~\cite{HIGG-2013-13}. 
Within the signal regions, for $W\gamma$ the corresponding uncertainties are 9\%, 53\%, and 100\% for $\ZeroJet$, $\OneJet$, and $\TwoJet$, respectively.
For $W\gamma^{\ast}$ they are 7\%, 30\%, and 26\%.  For the $\ZeroOneJetSimple$ signal regions, 
identical uncertainties apply in the SR and in the same-sign $VV$ CR for these processes. This 
results in a strong cancellation of the uncertainties in the predicted yields in the signal regions.

For the VBF $H\to{WW^{\ast}}$ contribution to the signal region, the cross-section uncertainties in 
the QCD scale (between $+2.6\%$ and $-2.8\%$) and PDF ($\pm 0.2\%$) are included~\cite{YR3}.
These have a negligible effect on the analysis, so additional uncertainties
in the VBF acceptance in the ggF phase space are not considered.

\section{Theory predictions}
\label{sec:theopred}
The results of the fiducial cross-section measurements are compared to analytical predictions calculated at parton level and to predictions by MC event generators at particle level.
An overview of the ggF predictions used is given in Table~\ref{tab:theopred}. All predictions are for $m_{H} = 125.0 \gev$ 
and $\rts = 8 \tev$, and use the CT10 PDF set unless stated otherwise. The values of the predictions are shown together with the results of the measurement in the following section.

\begin{table}[t!]
  \centering
  \caption{
   Summary of the ggF predictions used in comparison with the measured fiducial cross sections. The right column states the accuracy of each prediction in QCD.
  }
  \begin{tabular}{ll}
  \dbline
  \multicolumn{2}{c}{\bf Total cross-section predictions} \\
  LHC-XS~\cite{LHCHiggsCrossSectionWorkingGroup:2011ti}   & NNLO+NNLL \\
  \sgline
    \multicolumn{2}{c}{\bf Differential cross-section predictions} \\
  JetVHeto~\cite{Banfi:2012jm,Banfi:2012yh,Banfi:2013eda} & NNLO+NNLL \\
  ST~\cite{Stewart:2011cf} 	    & NNLO \\
  BLPTW~\cite{Boughezal:2013oha}   & NNLO+NNLL \\
  STWZ~\cite{Stewart:2013faa}  & NNLO+NNLL$^{\prime}$ \\
  N$^{3}$LO+NNLL+LL\_R~\cite{Banfi:2015pju}  & N$^{3}$LO+NNLL+LL\_R \\
  \sgline
  \multicolumn{2}{c}{\bf Monte Carlo event generators} \\
  {\POWHEG} {\NNLOPS}~\cite{Hamilton:2013fea,Hamilton:2015nsa} &  NNLO$_{\geq0j}$, NLO$_{\geq1j}$ \\
  \SHERPA 2.1.1~\cite{Gleisberg:2008ta,Krauss:2001iv,Hoeche:2012yf,Gehrmann:2012yg,Hoeche:2014lxa} 	& $H+0$, 1, 2 jets @NLO \\
  \MGFive~\cite{mg5aMCatNLO,Artoisenet:2013puc,Frederix:2012ps} & $H+0$, 1, 2 jets @NLO \\
    \dbline
  \end{tabular}
  \label{tab:theopred}
\end{table} 

The default prediction for the cross section of ggF Higgs boson production follows the recommendation of the LHC Higgs cross section working group~(LHC-XS) as introduced in Section~\ref{sec:mcpred}.
The \hwwenmn\ decay is included in the calculations and MC, with a branching fraction of 0.25\%.

For the efficiency $\varepsilon_{0}$ of the jet veto, a parton-level prediction is calculated at NNLO+NNLL accuracy by JetVHeto~\cite{Banfi:2012jm,Banfi:2012yh,Banfi:2013eda}. The uncertainty is taken as the maximum effect of the scale variations on the calculation, or the maximum deviation of the other calculations of $\varepsilon_{0}$ that differ by higher-order terms. 
An alternative prediction for $\varepsilon_{0}$ is given by the STWZ calculation~\cite{Stewart:2013faa}. The calculation has NNLO accuracy and is matched to a resummation at NNLL that accounts for the correct boundary conditions for the next-to-next-to-next-to-leading-logarithm resummation~(NNLL$^{\prime}$). 
This calculation also predicts the spectrum of $\pTj$. Another parton-level prediction of $\varepsilon_{0}$ follows the Stewart--Tackmann~(ST) prescription~\cite{Stewart:2011cf} utilising the total inclusive ggF cross section at NNLO accuracy in QCD and the inclusive $H+1$-jet cross section at NLO accuracy, calculated with HNNLO~\cite{Catani:2007vq,Grazzini:2008tf,Grazzini:2013mca}.  
Recently, a prediction for $\varepsilon_{0}$ has become available at N$^{3}$LO+NNLL accuracy with small-R resummation (LL\_R)~\cite{Banfi:2015pju}.
A parton-level prediction for the $\Njet$ distribution is given by the BLPTW method~\cite{Boughezal:2013oha}, combining the NNLO+NNLL-accurate inclusive and the NLO+NLL-accurate inclusive $H+1$-jet cross sections, including resummation in the covariance matrix. 

For comparisons to data, all parton-level predictions are corrected to particle level using the acceptance of the fiducial region and non-perturbative correction factors to account for the impact of hadronisation and underlying-event activity. These factors are determined using {\POWHEG} {\NNLOPS}+\PythiaEight~\cite{Hamilton:2013fea,Hamilton:2015nsa} with the associated uncertainties from the renormalisation and factorisation scales as well as the PDFs. 
An uncertainty is assigned to the non-perturbative correction by comparing \PythiaEight with \HERWIG. The uncertainties applied are between 0.5\% and 7\%. All factors are given in HEPDATA.

Particle-level predictions for the measured differential cross sections are provided by MC event generators. The most precise prediction for inclusive ggF production is given by {\POWHEG} {\NNLOPS}, which is accurate to NNLO for the inclusive production and to NLO for the inclusive $H+1$-jet production, combining the \textsc{MINLO}~\cite{Hamilton:2012np} method with an NNLO calculation of the Higgs boson rapidity using HNNLO. Furthermore, it includes finite quark masses~\cite{Hamilton:2015nsa}. The sample is generated using the CT10nnlo PDF set~\cite{Gao:2013xoa} and is interfaced to \PythiaEight for parton showering.
The uncertainties include a 27-point QCD scale variation described in Ref.~\cite{Hamilton:2013fea}, as well as a PDF uncertainty, obtained from variations of the CT10 PDF set.

Another ggF MC prediction is generated with \SHERPA (v.2.1.1)~\cite{Gleisberg:2008ta,Krauss:2001iv}. Here, the inclusive Higgs boson, inclusive $H+1$-jet, and inclusive $H+2$-jets production cross sections are calculated at NLO accuracy. The $H+2$-jets matrix elements are generated via an MCFM interface within \SHERPA. These calculations are combined using the \textsc{MEPS@NLO} method~\cite{Hoeche:2012yf,Gehrmann:2012yg}. 
The factorisation, renormalisation, resummation, and merging scales are varied to determine an uncertainty as described in Ref.~\cite{Hoeche:2014lxa}. Additionally, the variations of the CT10 PDF set are included.

A similar NLO-merged $H$+ (0, 1, 2)-jets sample is generated with {\MGFive} (v.2.3.2.2)~\cite{mg5aMCatNLO,Artoisenet:2013puc} where the different calculations are combined using the \textsc{FxFx} scheme~\cite{Frederix:2012ps}. {\MGFive} is interfaced to \PythiaEight for parton showering. 
Variations of the factorisation, renormalisation, and merging scales, and of the CT10 PDF set, are evaluated for each prediction. The differences in the predictions are taken as uncertainties.

\section{Results}
\label{sec:results}
The cross section of ggF Higgs boson production in the fiducial region defined in Table~\ref{tab:fidsel} is measured to be:
\begin{linenomath}
\begin{align*}
\sigma_{\mathrm{ggF}}^{\mathrm{fid}} &= 36.0 \pm 7.2\mathrm{(stat)} \pm 6.4\mathrm{(sys)} \pm 1.0\mathrm{(lumi)}\; \mathrm{fb} \\
							&=  36.0 \pm 9.7\; \mathrm{fb} 
\end{align*}
\end{linenomath}
where (stat) includes all statistical uncertainties from the signal and control regions, and (sys) refers to the sum 
in quadrature of the experimental and theoretical systematic 
uncertainties.  The mass of the Higgs boson is assumed to be $m_{H} = 125.0 \gev$.
The fiducial cross section is calculated from the number of events after the event selection and detector corrections, using an 
integrated luminosity of 20.3~\ifb\ with an associated uncertainty of 2.8\%. This is derived 
following the same methodology as in Ref.~\cite{DAPR-2011-01}.
More details of the sources of systematic uncertainty are given in Table~\ref{tab:total}.
The uncertainty categories used in this and all tables in this section are as follows.  Statistical
uncertainties are quoted separately for the signal region data, the control region data, and the MC simulated events.
Experimental uncertainties (``Exp.'') are grouped according to the reconstructed object they effect. The ``Exp.~other'' 
category includes uncertainties in the modelling of pile-up events, electrons from conversions, and the modelling of
the $\pT$ of $Z$ bosons with $\ZeroJet$.  Theory uncertainties are grouped by process, with the subdominant background
uncertainties collected in the ``Theory other backgrounds'' line.  The ``Detector corrections'' line gives the effect of the
use of the ggF signal MC sample to construct the migration matrix, as described in Section~\ref{sec:unfunc}.
\begin{table}[tbp!]
\centering
\caption{\label{tab:total} Relative uncertainties (in \%) in the measured total fiducial cross section. }
\begin{tabular}{lr} 
\dbline 
Source & $\Delta\sigma_{\mathrm{ggF}}^{\mathrm{fid}}/\sigma_{\mathrm{ggF}}^{\mathrm{fid}}$ $[\%]$  \\ 
\sgline
SR data statistical		& 17\phantom{.3}\\
MC statistical	& 3.0 \\
CR data statistical		& 9.9 \\
Exp. JER & 4.9 \\
Exp. JES & 2.1 \\
Exp. $b$-tag  & 3.3 \\
Exp. leptons 	&  5.5  \\
Exp. $\MET$	&  2.2  \\
Exp. other	&  4.2\\
Theory ($WW$)	& 14\phantom{.2}  \\
Theory (top)	& 7.1\\
Theory (other backgrounds) & 5.6\\
Theory (signal) & 2.5 \\
Detector corrections	& 0.4 \\
\sgline
Total		& 27\phantom{.3} \\
\dbline
\end{tabular}
\end{table}

The prediction of the fiducial cross section is given by the LHC-XS calculation as 
\begin{linenomath}
$$\setlength\arraycolsep{0.1em}
 \begin{array}{rclclcl}
 
  \textrm{LHC-XS:} & \sigma_{\mathrm{ggF}}^{\mathrm{fid}} &=& 25.1 ^{+1.8}_{-2.0} \mathrm{(QCD scales)} ^{+1.9}_{-1.7} \mathrm{(PDF)} \; \mathrm{fb} &=& 25.1 \pm 2.6\; \mathrm{fb.} \\
 \end{array}
 $$
 \end{linenomath}
Reference~\cite{HIGG-2013-13} also reports ggF fiducial cross sections for events with $\ZeroJet$ and $\OneJet$, but 
with modified fiducial region selections, among which the most important one is a lower threshold of $10 \GeV$ on the subleading lepton $\pT$.
The ratio of the observed to predicted SM cross sections in that analysis is statistically compatible with the results shown here.

The dependence of the cross-section measurement on $\mH$ is mainly due to acceptance effects
and is approximated by a linear function, which is sufficient within the experimental uncertainties in the Higgs boson mass~\cite{HIGG-2014-14}. 
The function is determined using dedicated signal samples with different values of $\mH$
and has a slope of $- 0.20\; \mathrm{fb}/\GeV$.

\subsection{Differential fiducial cross sections}

Differential fiducial cross sections are measured in bins of the $\Njet$, $\pTH$, $\yll$ and $\pTj$ distributions. 
For the $\pTH$, $\yll$ and $\pTj$ distributions, the cross sections are measured in separate bins of $\Njet$ to fully take correlations 
into account between the different $\Njet$ categories and the variable itself. After detector corrections the 
distributions are integrated over $\Njet$, and the uncertainties are combined accounting for correlations.
The measured differential 
fiducial cross sections as a function of $\Njet$, $\pTH$, $\yll$, and $\pTj$ are given in Tables~\ref{tab:njets}--\ref{tab:pTj1}, together 
with a summary of the associated uncertainties. 
The dominant systematic uncertainties are in the background model, in particular the Monte-Carlo modelling of 
  top-quark and $WW$ backgrounds, which are evaluated as described in Section~\ref{sec:bgsys}. 
 The large background fraction in the signal region amplifies the effect of even small (from about $1$\% to $5$\%) 
   residual extrapolation uncertainties after normalisation in a control region.
  The uncertainties from the experimental inputs are also non-negligible. 

Figure~\ref{fig:all_observed} shows the measured
differential cross sections as a function of $\Njet$, $\pTH$, $\yll$, and $\pTj$.
The results are compared to particle-level predictions from {\POWHEG} {\NNLOPS}, \SHERPA, and {\MGFive} for ggF Higgs boson production.  The predictions 
are generated as described in Section~\ref{sec:theopred} and normalised to the cross-section predictions calculated according to the
prescription from the LHC-XS working group. In addition, the results for the $\Njet$ distribution are compared to the parton-level 
BLPTW calculation, and the results for the $\pTj$ distribution are compared to the parton-level STWZ calculation.
The ratios of the results to the predictions are given in the lower panel of each figure. 
The measured distributions agree with the predictions within the uncertainties, except for $\yll$, where the 
data have a more central mean $\yll$ than the predictions.  The statistical and systematic uncertainties are comparable for most bins. 

\begin{table}[tbp!]
\centering
\caption{\label{tab:njets} Measured and predicted fiducial cross section in fb as a 
  function of $\Njet$. 
  Predicted values are from {\POWHEG} {\NNLOPS}+\PythiaEight, normalised to the LHC-XS working group recommended
  cross section, as described in Section~\ref{sec:theopred}.  Total uncertainties in the measurement are
  given along with their relative composition in terms of source.
} 
\begin{tabular}{lrrr} 
\dbline
$\Njet$ &  0 & 1 & $\geq 2$  \\ 
\sgline
d$\sigma/\mathrm{d}\Njet$~[fb]	& 19.0 & 8.2 & 8.8\\
Statistical uncertainty		&  4.5 & 3.5 & 5.0 \\ 
Total uncertainty	 	&  6.8 & 4.0 & 5.9 \\
\sgline
Predicted~d$\sigma/\mathrm{d}\Njet$~[fb] (NNLOPS)	& 14.7  & 7.0  & 3.4	\\
Uncertainty in prediction	&  1.8  & 0.9  & 0.6 \\
\sgline
SR data statistical	& 20\% & 38\% & 54\% \\
MC statistical		&  4\% &  7\% &  9\% \\
CR data statistical	& 12\% & 18\% & 14\% \\
Exp. JER & 5\% & 4\% & 7\% \\
Exp. JES & 1\% & 10\% & 6\% \\
Exp. $b$-tag  & 1\% & 4\% & 8\% \\
Exp. leptons 	&  6\% & 6\% & 6\% \\
Exp. $\MET$	&  2\% & 4\% & 4\% \\
Exp. other	&  5\% & 4\% & 3\% \\
Theory ($WW$)	& 24\% & 15\% & 5\% \\
Theory (top)	& 2\% & 4\% & 24\% \\
Theory (other backgrounds) & 5\% & 6\% & 21\% \\
Theory (signal)	&  4\% & 6\% & 3\% \\
Detector corrections	&  <1\% &  4\% &  5\% \\
\sgline
Total uncertainty		&36\% & 48\% & 67\% \\
\dbline
\end{tabular}
\end{table}

\begin{table}[tbp!]
\centering
\caption{\label{tab:pTH} Measured and predicted differential fiducial cross section in fb/$\GeV$ as a function of $\pTH$. 
  Predicted values are from {\POWHEG} {\NNLOPS}+\PythiaEight, normalised to the LHC-XS working group recommended
  cross section, as described in Section~\ref{sec:theopred}.  Total uncertainties in the measurement are
  given along with their relative composition in terms of source.
} 
\begin{tabular}{lrrr} \dbline 
$\pTH$~$[\GeV]$ & $[0, 20]$ & $[20,60]$ & $[60,300]$ \\ 
\sgline
d$\sigma/\mathrm{d}\pTH$~[fb/$\GeV$] & 0.61 & 0.39 & 0.034\\
Statistical uncertainty 	& 0.16 & 0.09 & 0.021 \\
Total uncertainty		& 0.29 & 0.15 & 0.027 \\
\sgline
Predicted~d$\sigma/\mathrm{d}\pTH$~[fb/$\GeV$] (NNLOPS)	& 0.48 	& 0.25	& 0.022		\\
Uncertainty in prediction		& 0.05	& 0.03	& 0.005		\\
     \sgline
SR data statistical 	&22\% & 22\% & 60\% \\
MC statistical 		&4\% & 4\% & 10\% \\
CR data statistical 	&13\% & 5\% & 18\% \\
Exp. JER & 7\% & 4\% & 16\% \\
Exp. JES & 6\% & 10\% & 17\% \\
Exp. $b$-tag  & 2\% & 4\% & 8\% \\
Exp. leptons 		&  7\% & 6\% & 7\% \\
Exp. $\MET$		&  9\% & 8\% & 7\% \\
Exp. other		&  7\% & 4\% & 4\% \\
Theory ($WW$)		& 31\% & 17\% & 13\% \\
Theory (top)		& 4\% & 7\% & 25\% \\
Theory (other backgrounds) 	& 6\% & 8\% & 14\% \\
Theory (signal) 	&14\% & 1\% & 6\% \\
Detector corrections 	&<1\% & 3\% & 3\% \\
\sgline
Total 			&47\% & 37\% & 77\% \\
\dbline
\end{tabular}
\end{table}

\begin{table}[tbp!]
\centering
\caption{\label{tab:Yll} Measured and predicted differential fiducial cross section in fb per unit rapidity as a function of $\yll$.
  Predicted values are from {\POWHEG} {\NNLOPS}+\PythiaEight, normalised to the LHC-XS working group recommended
  cross section, as described in Section~\ref{sec:theopred}.  Total uncertainties in the measurement are
  given along with their relative composition in terms of source.
} 
\begin{tabular}{lrrr} 
\dbline 
$\yll$ & $[0.0, 0.6]$ & $[0.6,1.2]$ & $[1.2,2.5]$ \\ 
\sgline
d$\sigma/\mathrm{d}\yll$~[fb] 	& 31\phantom{.0} & 9.5 & 9.5\\
Statistical uncertainty	& 7.3 & 5.0 & 3.5 \\
Total uncertainty		& 10\phantom{.0} & 6.5 & 5.2 \\
\sgline
Predicted~d$\sigma/\mathrm{d}\yll$~[fb] (NNLOPS)	& 15.9	& 13.0	& 5.9	\\ 
Uncertainty in prediction		&  1.7	&  1.4	& 0.6 	\\
\sgline
SR data statistical 	&22\% & 52\% & 33\% \\
MC statistical 		&3\% & 9\% & 6\% \\
CR data statistical	&9\% & 1\% & 16\% \\
Exp. JER & 4\% & 10\% & 4\% \\
Exp. JES & 5\% & 9\% & 6\% \\
Exp. $b$-tag  & 3\% & 4\% & 5\% \\
Exp. leptons 	&  4\% & 10\% & 9\% \\
Exp. $\MET$	&  3\% & 8\% & 4\% \\
Exp. other	&  4\% & 8\% & 6\% \\
Theory ($WW$)	& 15\% & 31\% & 20\% \\
Theory (top)	& 12\% & 14\% & 8\% \\
Theory (other backgrounds) & 3\% & 7\% & 17\% \\
Theory (signal)	&4\% & 6\% & 3\% \\
Detector corrections	&<1\% & <1\% & 1\% \\
\sgline
Total		&33\% & 69\% & 53\% \\
\dbline
\end{tabular}
\end{table}

\begin{table}[tbp!]
\centering
\caption{\label{tab:pTj1} Measured and predicted differential fiducial cross section in fb/$\GeV$ as a function of $\pTj$. 
  Predicted values are from {\POWHEG} {\NNLOPS}+\PythiaEight, normalised to the LHC-XS working group recommended
  cross section, as described in Section~\ref{sec:theopred}.  Total uncertainties in the measurement are
  given along with their relative composition in terms of source.
} 
\begin{tabular}{lrrr} 
\dbline 
$\pTj$~[$\GeV$] & $[0, 30]$ & $[30,60]$ & $[60,300]$ \\ 
\sgline
$ \mathrm{d}\sigma/\mathrm{d}\pTj$~[fb/$\GeV$] 	& 0.69 & 0.26 & 0.034\\
Statistical uncertainty 		& 0.16 & 0.10 & 0.021 \\
Total uncertainty   	& 0.24 & 0.13 & 0.025 \\
\sgline
Predicted~$\mathrm{d}\sigma/\mathrm{d}\pTj$~[fb/$\GeV$] (NNLOPS)	& 0.53 	& 0.17	& 0.016		\\
Uncertainty in prediction		& 0.06  & 0.02	& 0.004		\\
\sgline
SR data statistical	&19\% & 40\% & 61\% \\
MC statistical	&3\% & 7\% & 10\% \\
CR data statistical 	&12\% & 2\% & 18\% \\
Exp. JER & 4\% & 6\% & 10\% \\
Exp. JES & 2\% & 14\% & 15\% \\
Exp. $b$-tag  & 1\% & 8\% & 10\% \\
Exp. leptons 	&  6\% & 6\% & 8\% \\
Exp. $\MET$	&  2\% & 6\% & 4\% \\
Exp. other	&  5\% & 5\% & 4\% \\
Theory ($WW$)	& 23\% & 12\% & 14\% \\
Theory (top)	& 2\% & 13\% & 23\% \\
Theory (other backgrounds) & 5\% & 13\% & 13\% \\
Theory (signal)	&5\% & 4\% & 3\% \\
Detector corrections	&<1\% & <1\% & <1\% \\
\sgline
Total 		&34\% & 51\% & 75\% \\
\dbline
\end{tabular}
\end{table}

\begin{figure}[!tbh]
  \centering
  \subfloat[$\Njet$]{
    \includegraphics[width=0.46\textwidth]{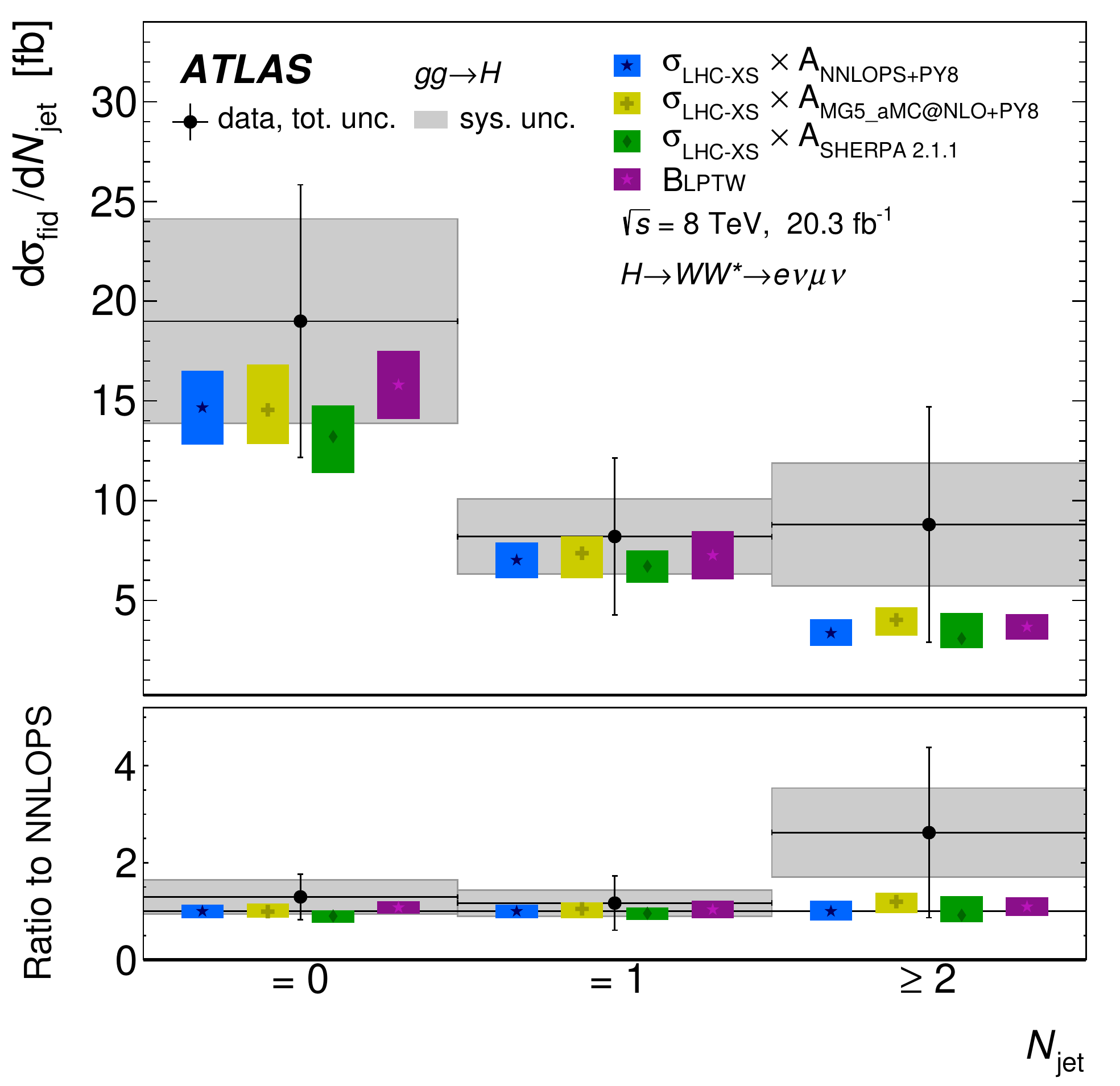}
  }
  \subfloat[$\pTH$]{
    \includegraphics[width=0.46\textwidth]{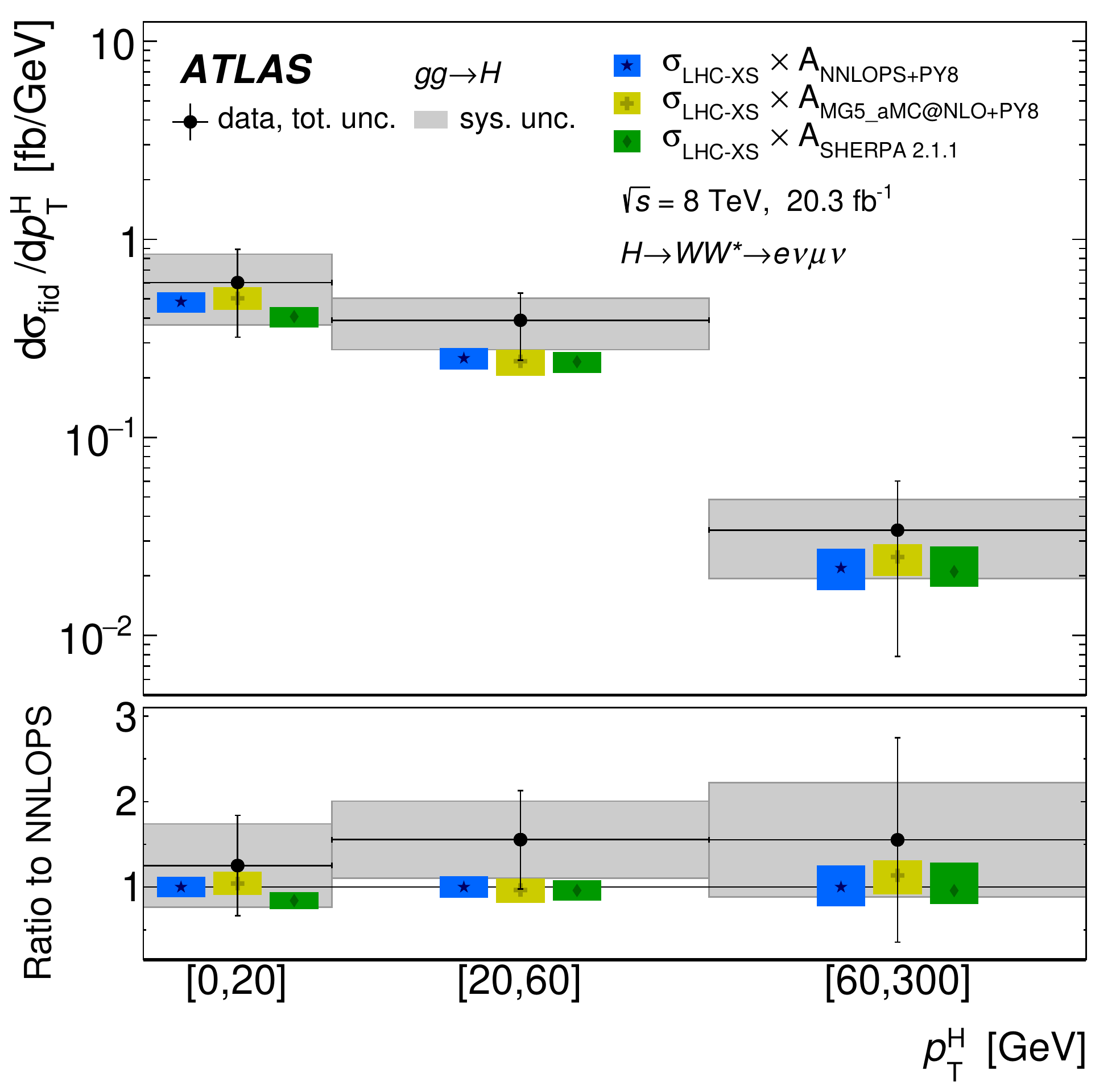}
  }
\newline
  \subfloat[$\yll$]{
    \includegraphics[width=0.46\textwidth]{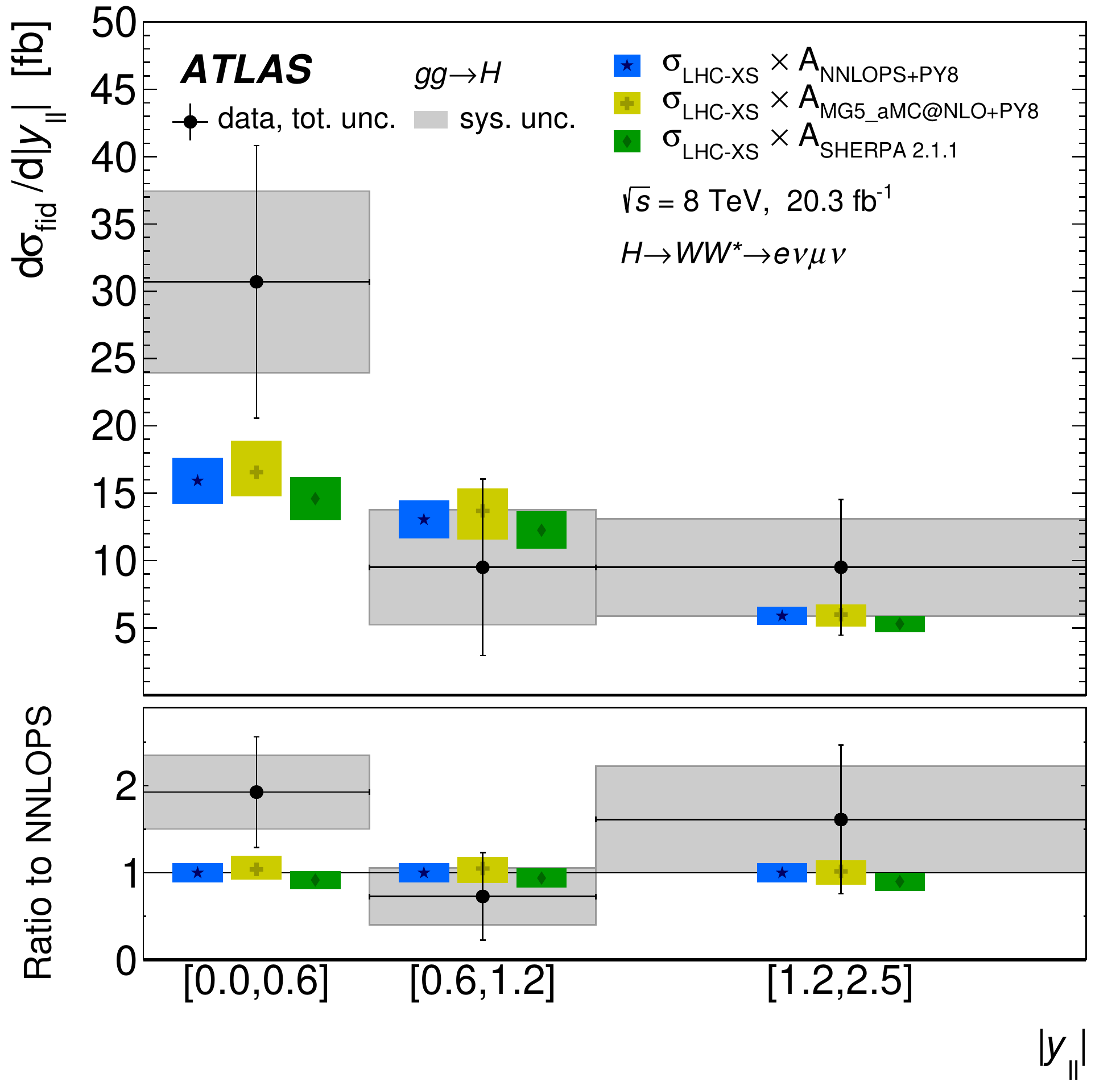}
  }
  \subfloat[$\pTj$]{
    \includegraphics[width=0.46\textwidth]{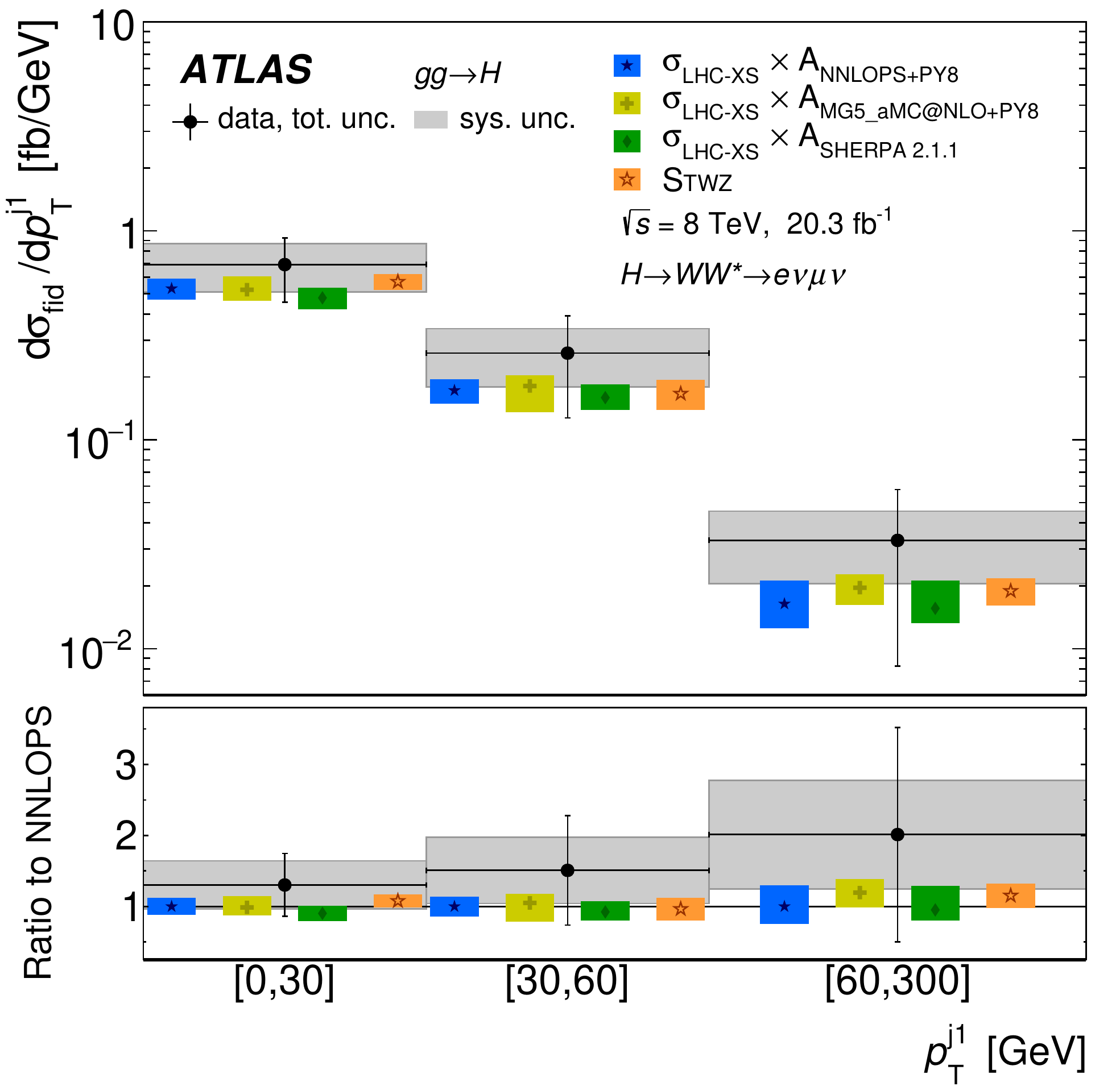}
  }
  { \caption{Measured fiducial differential cross section as a function of (a) $\Njet$, (b) $\pTH$, (c) $\yll$, and (d) $\pTj$, overlaid with the 
	      signal predictions.  The $[0,30] \gev$ bin of the $\pTj$ distribution includes events with no reconstructed jets. 
	      The systematic uncertainty at each point is shown by a grey band labelled ``sys. unc.''
	     and includes the experimental and theoretical uncertainties.  The uncertainty bar, labelled ``data, tot. unc.'' is the total 
	      uncertainty and includes all systematic and statistical uncertainties. The measured results are compared to various theoretical predictions.
  \label{fig:all_observed}
    }}
\end{figure}

\subsection{Normalised differential fiducial cross sections}\label{sec:results_norm}

To reduce the impact of systematic uncertainties, normalised differential cross sections $1/\sigma \cdot (\mathrm{d}\sigma/\mathrm{d}X_{i})$
are calculated by dividing the differential cross section by the total fiducial cross section evaluated by integrating over all bins of variable $X$.
The normalised differential cross sections as functions of $\Njet$, $\pTH$, $\yll$, and $\pTj$ are given in 
Tables~\ref{tab:njets_norm}--\ref{tab:pTj1_norm}, along with details of the associated uncertainties.
The distributions are shown in Figure~\ref{fig:all_observed_norm} compared to particle-level 
predictions of ggF Higgs boson production by {\POWHEG} {\NNLOPS}, \SHERPA, and {\MGFive} that are generated as described in Section~\ref{sec:theopred}.
In each figure, the ratio of the result to the predictions is shown below the distribution.
The reduced uncertainties result in a more stringent comparison of the measured and predicted distributions. 
The level of agreement is still good although the trend in $\yll$ is enhanced and a slight trend towards
higher $\Njet$ and $\pTj$ appears in the data.  

\begin{table}[tbp!]
\centering
\caption{\label{tab:njets_norm} Measured and predicted normalised differential fiducial cross section as a function of $\Njet$.
  Predicted values are from {\POWHEG} {\NNLOPS}+\PythiaEight, normalised to the LHC-XS working group recommended
  cross section, as described in Section~\ref{sec:theopred}.  Total uncertainties in the measurement are
  given along with their relative composition in terms of source.
} 
\begin{tabular}{lrrr} \dbline 
$\Njet$ 			&  0 & 1 & $\geq 2$  \\ 
\sgline
$ 1/\sigma$~$\mathrm{d}\sigma/\mathrm{d}\Njet$  & 0.53 & 0.23 & 0.24\\
Statistical uncertainty	& 0.11 & 0.09 & 0.12 \\
Total uncertainty	& 0.14 & 0.10 & 0.14 \\
\sgline
Predicted~$1/\sigma$~$\mathrm{d}\sigma/\mathrm{d}\Njet$ (NNLOPS)	& 0.59	& 0.28	& 0.13	\\
Uncertainty in prediction			& 0.04	& 0.02	& 0.02	\\
\sgline
SR data statistical 	&19\% & 34\% & 42\% \\
MC statistical 	& 4\% & 8\% & 17\% \\
CR data statistical 	& 9\% & 16\% & 14\% \\
Exp. JER & <1\% & 1\% & 2\% \\
Exp. JES & 3\% & 7\% & 4\% \\
Exp. $b$-tag  & 3\% & 3\% & 5\% \\
Exp. leptons  &  2\% & 2\% & 4\% \\
Exp. $\MET$  &  1\% & 4\% & 4\% \\
Exp. other  &  2\% & 2\% & 3\% \\
Theory ($WW$) & 12\% & 15\% & 17\% \\
Theory (top)  & 7\% & 5\% & 18\% \\
Theory (other backgrounds) & 6\% & 5\% & 16\% \\
Theory (signal) 	& 1\% & 3\% & 5\% \\
Detector corrections 	& <1\% & 4\% & 4\% \\
\sgline
Total &26\% & 43\% & 57\% \\
\dbline
\end{tabular}
\end{table}

\begin{table}[tbp!]
\centering
\caption{\label{tab:pTH_norm} Measured and predicted normalised differential fiducial cross section as a function of $\pTH$. 
  Predicted values are from {\POWHEG} {\NNLOPS}+\PythiaEight, normalised to the LHC-XS working group recommended
  cross section, as described in Section~\ref{sec:theopred}.  Total uncertainties in the measurement are
  given along with their relative composition in terms of source.
} 
\begin{tabular}{lrrr} 
\dbline 
$\pTH$~[$\GeV$] & $[0, 20]$ & $[20,60]$ & $[60,300]$ \\ 
\sgline
$ 1/\sigma$~$\mathrm{d}\sigma/\mathrm{d}\pTH$~[$10^{-3}\GeV^{-1}$] & 17.0 & 11.0 & 0.96\\
Statistical uncertainty	& 3.5 & 2.0 & 0.50 \\
Total uncertainty	& 6.0 & 3.4 & 0.63 \\
\sgline
Predicted~$1/\sigma$~$\mathrm{d}\sigma/\mathrm{d}\pTH$~[$10^{-3}\GeV^{-1}$] (NNLOPS)	& 19.4 & 10.0 & 0.88 \\
 Uncertainty in prediction						& 0.7 & 0.5 & 0.2 \\
 \sgline
SR data statistical 	&20\% & 18\% & 48\% \\
MC statistical 	&4\% & 3\% & 8\% \\
CR data statistical 	&8\% & 7\% & 18\% \\
Exp. JER & 2\% & 4\% & 11\% \\
Exp. JES & 8\% & 9\% & 16\% \\
Exp. $b$-tag  & 4\% & 4\% & 6\% \\ 
Exp. leptons  &  3\% & 2\% & 5\% \\
Exp. $\MET$  &  10\% & 8\% & 7\% \\
Exp. other  &  4\% & 2\% & 4\% \\
Theory ($WW$) & 19\% & 15\% & 21\% \\
Theory (top)  & 9\% & 8\% & 17\% \\
Theory (other backgrounds) & 7\% & 8\% & 12\% \\
Theory (signal) 	&10\% & 2\% & 10\% \\
Detector corrections 	&<1\% & 3\% & 3\% \\
\sgline
Total 		 &37\% & 31\% & 65\% \\
\dbline
\end{tabular}
\end{table}

\begin{table}[tbp!]
\centering
\caption{\label{tab:Yll_norm} Measured and predicted normalised differential fiducial cross section as a function of $\yll$. 
  Predicted values are from {\POWHEG} {\NNLOPS}+\PythiaEight, normalised to the LHC-XS working group recommended
  cross section, as described in Section~\ref{sec:theopred}.  Total uncertainties in the measurement are
  given along with their relative composition in terms of source.
} 
\begin{tabular}{lrrr}
\dbline 
$|y(\ell\ell)|$ & $[0.0, 0.6]$ & $[0.6,1.2]$ & $[1.2,2.5]$ \\
 \sgline
$ 1/\sigma$~$\mathrm{d}\sigma/\mathrm{d}|y(\ell\ell)|$ & 0.83 & 0.27 & 0.26\\
Statistical uncertainty & 0.17 & 0.13 & 0.08 \\
Total uncertainty		 & 0.22 & 0.15 & 0.11 \\
 \sgline
 Predicted$1/\sigma$~$\mathrm{d}\sigma/\mathrm{d}|y(\ell\ell)|$ (NNLOPS)	& 0.636 & 0.521 & 0.235 \\
 Uncertainty in prediction		& 0.004 & 0.001 & 0.004 \\
\sgline
SR data statistical	&18\% & 48\% & 26\% \\
MC statistical	& 3\% & 8\% & 5\% \\
CR data statistical	& 7\% & 6\% & 14\% \\
Exp. JER & 2\% & 5\% & 2\% \\
Exp. JES & 4\% & 9\% & 7\% \\
Exp. $b$-tag  & 3\% & 5\% & 5\% \\
Exp. leptons  &  3\% & 5\% & 5\% \\
Exp. $\MET$  &  3\% & 7\% & 4\% \\
Exp. other  &  3\% & 6\% & 5\% \\
Theory ($WW$) & 11\% & 21\% & 18\% \\
Theory (top)  & 10\% & 15\% & 9\% \\
Theory (other backgrounds) & 5\% & 8\% & 17\% \\
Theory (signal)	& <1\% & 2\% & 1\% \\
Detector corrections	& <1\% & <1\% & <1\% \\
\sgline
Total		&27\% & 60\% & 43\%  \\
\dbline
\end{tabular}
\end{table}

\begin{table}[tbp!]
\centering
\caption{\label{tab:pTj1_norm} Measured and predicted normalised differential cross section as a function of $\pTj$. 
  Predicted values are from {\POWHEG} {\NNLOPS}+\PythiaEight, normalised to the LHC-XS working group recommended
  cross section, as described in Section~\ref{sec:theopred}.  Total uncertainties in the measurement are
  given along with their relative composition in terms of source.
} 
\begin{tabular}{lrrr}
\dbline 
$\pTj$~[$\GeV$] & $[0, 30]$ & $[30,60]$ & $[60,300]$ \\
\sgline
$ 1/\sigma$~$\mathrm{d}\sigma/\mathrm{d}\pTj$~[$10^{-3}\GeV^{-1}$] & 19.0 & 7.0 & 0.91\\
Statistical uncertainty & 3.7 & 2.7 & 0.51 \\
Total uncertainty		 & 4.7 & 3.3 & 0.58 \\
 \sgline
 Predicted~$1/\sigma~\mathrm{d}\sigma/\mathrm{d}\pTj$~[$10^{-3}\GeV^{-1}$] (NNLOPS)	& 21.2	& 6.9	& 0.66	\\
 Uncertainty in prediction		& 0.7 	& 0.5	& 0.16	\\
\sgline
SR data statistical	 &17\% & 36\% & 49\%  \\
MC statistical	&3\% & 6\% & 9\% \\
CR data statistical	&7\% & 8\% & 18\% \\
Exp. JER & 2\% & 3\% & 5\% \\
Exp. JES & 3\% & 13\% & 14\% \\
Exp. $b$-tag  & 3\% & 7\% & 9\% \\
Exp. leptons  &  2\% & 3\% & 5\% \\
Exp. $\MET$  &  1\% & 6\% & 4\% \\
Exp. other  &  2\% & 3\% & 5\% \\
Theory ($WW$) & 11\% & 17\% & 17\% \\
Theory (top)  & 7\% & 9\% & 18\% \\
Theory (other backgrounds) & 5\% & 11\% & 11\% \\
Theory (signal)	&2\% & 2\% & 5\% \\
Detector corrections	&<1\% & <1\% & <1\% \\
\sgline
Total & 24\% & 47\% & 63\% \\
\dbline
\end{tabular}
\end{table}

\begin{figure}[!tbh]
  \centering
  \subfloat[$\Njet$]{
    \includegraphics[width=0.46\textwidth]{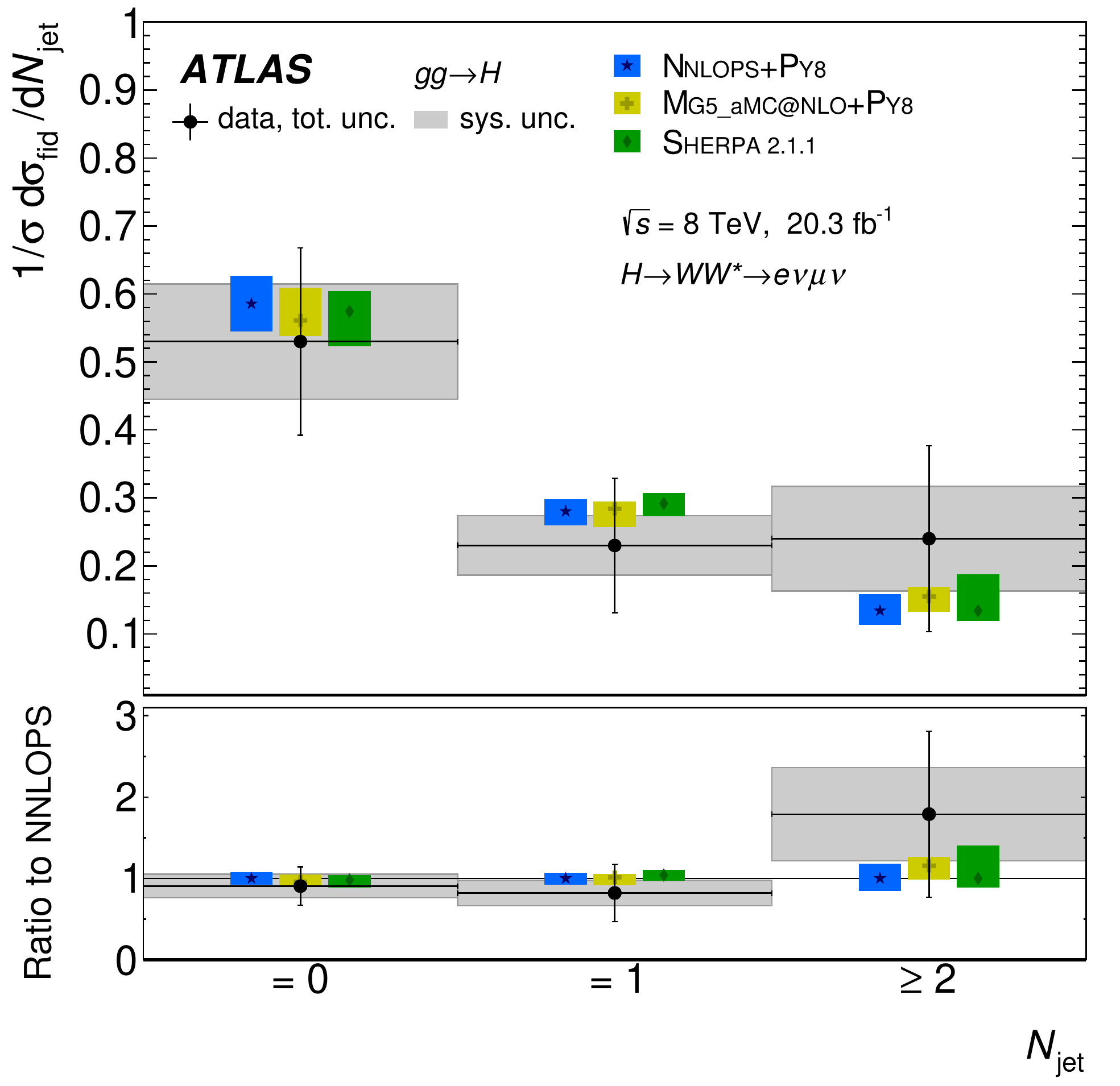}
  }
  \subfloat[$\pTH$]{
    \includegraphics[width=0.46\textwidth]{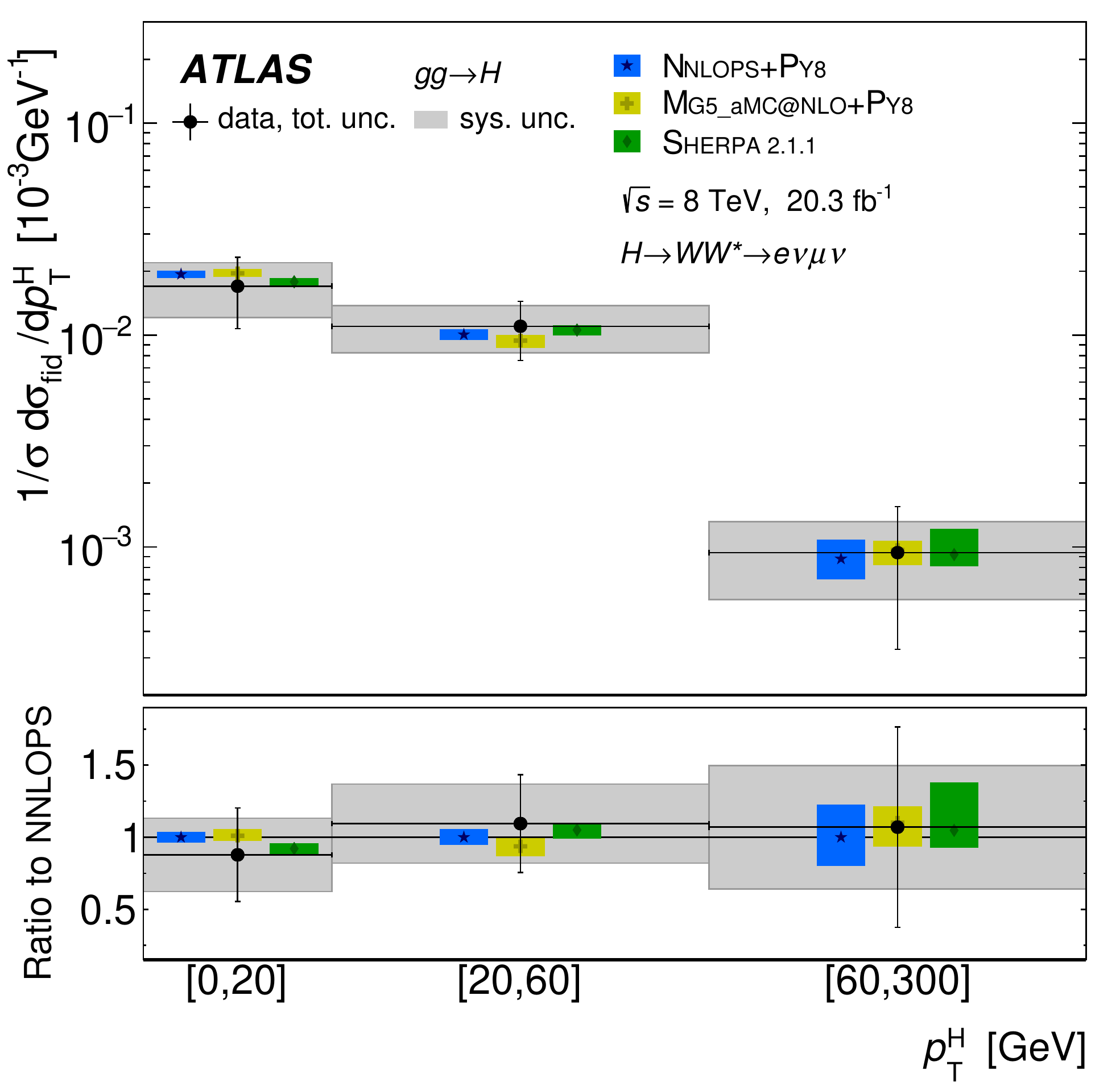}
  }
\newline
  \subfloat[$\yll$]{
    \includegraphics[width=0.46\textwidth]{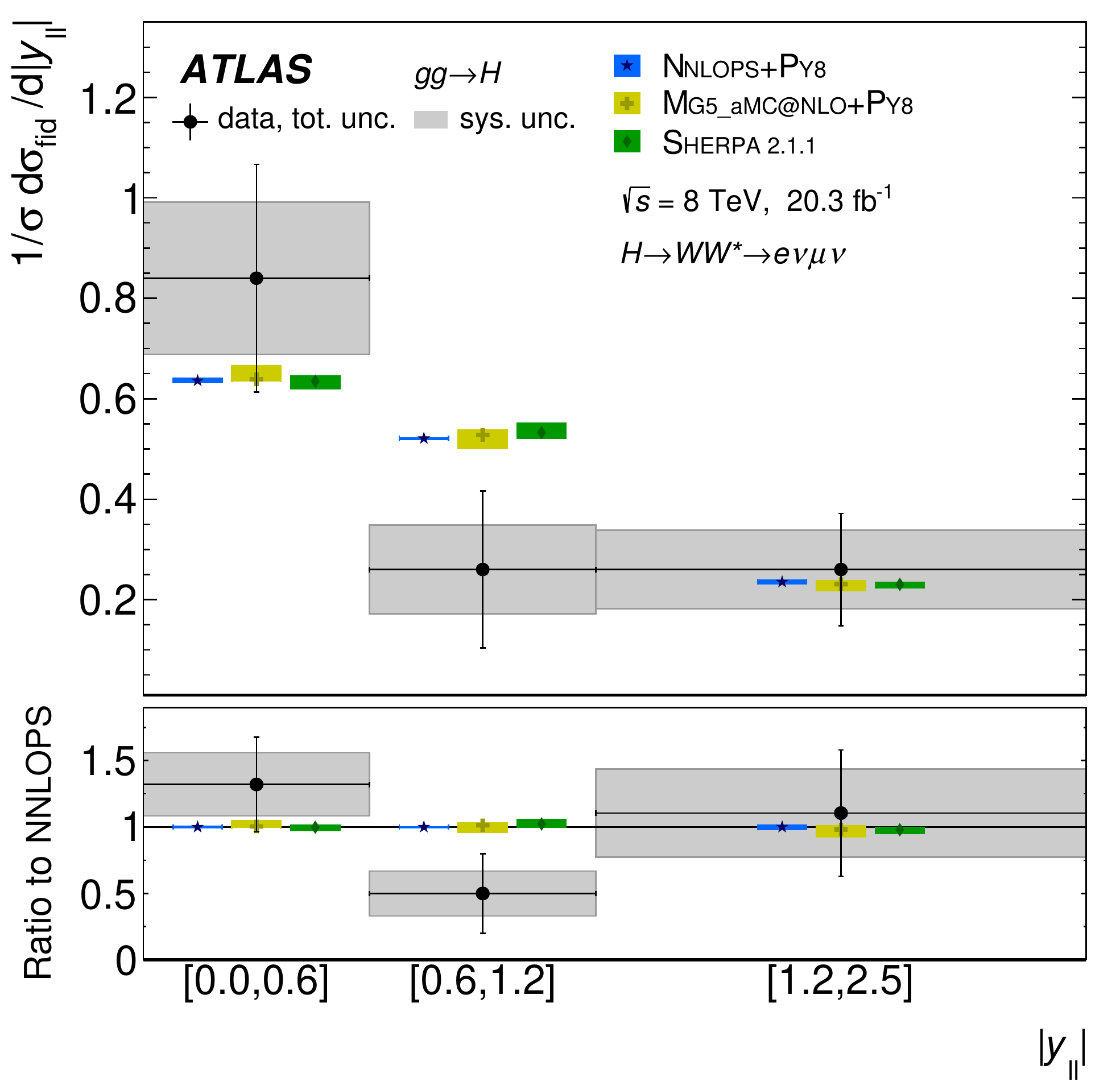}
  }
  \subfloat[$\pTj$]{
    \includegraphics[width=0.46\textwidth]{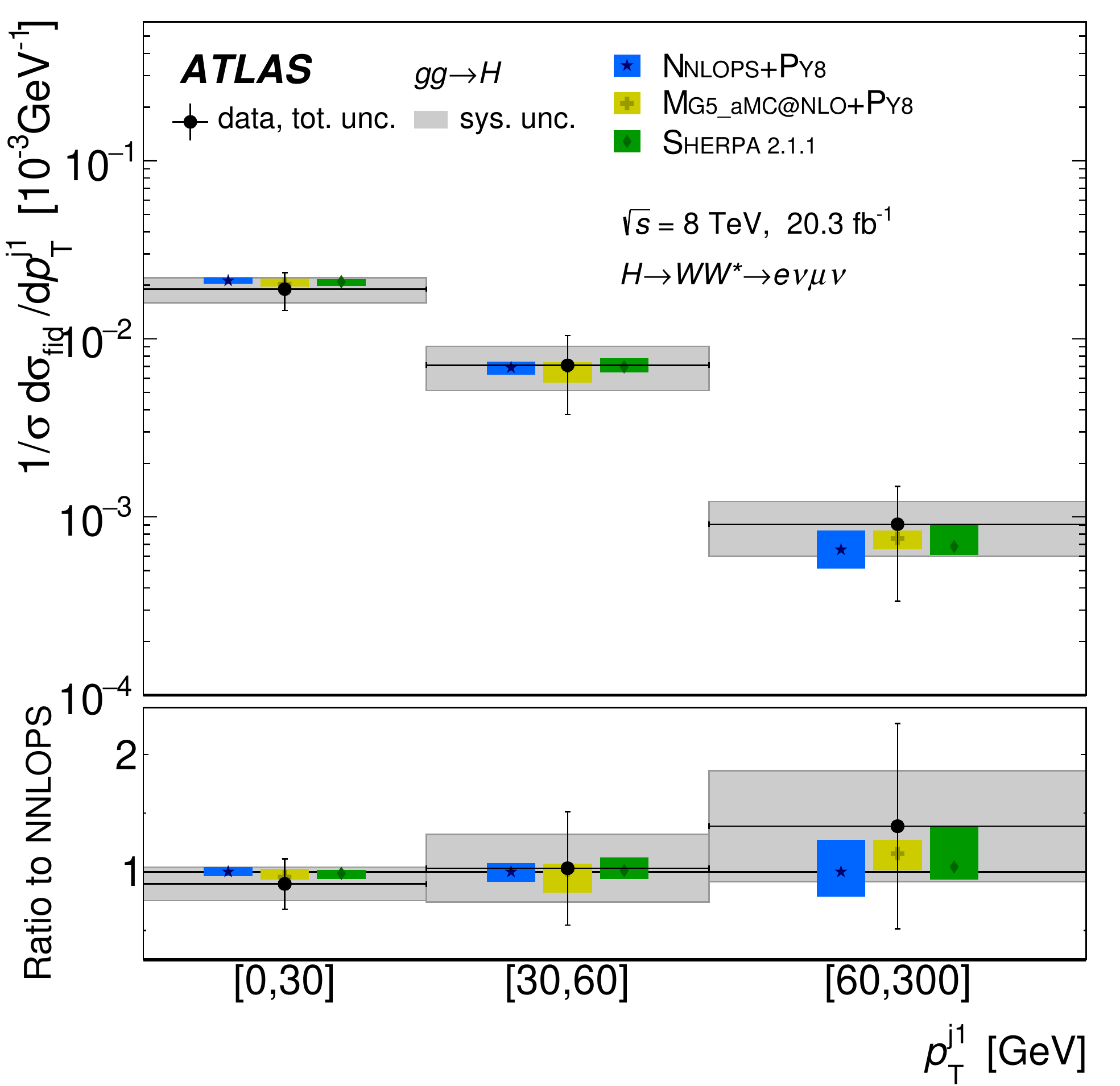}
  }
  { \caption{Normalised fiducial differential cross section measurements as a function of (a) $\Njet$, (b) $\pTH$, (c) $\yll$, and (d) $\pTj$, overlaid with the 
	      signal predictions.  The $[0,30] \gev$ bin of the $\pTj$ distribution includes events with no reconstructed jets. 
	      The systematic uncertainty at each point is shown by a grey band labelled ``sys. unc.''
	     and includes the experimental and theoretical uncertainties.  The uncertainty bar, labelled ``data, tot. unc.'' is the total 
	      uncertainty and includes all systematic and statistical uncertainties. The measured results are compared to various theoretical predictions.
  \label{fig:all_observed_norm}
    }}
\end{figure}

\subsection{Jet-veto efficiency}\label{sec:jve_results}

The jet-veto efficiency $\varepsilon_{0}$ for the $H+0$-jet events is defined at particle level as the fraction of events in the 
fiducial region with the leading particle-level jet below a given threshold.  
This is measured using the leading-jet $\pT$ distribution, since the
lowest-$\pT$ bin contains exactly the fraction of events with the leading jet
below the threshold of either $\pTj = 30\GeV$ or $\pTj = 40\GeV$.  The jet-veto efficiency for the jet 
selection used in the analysis, $25\GeV$ for central jets ($|\eta|<2.4$) and $30\GeV$ 
for forward jets ($2.4<|\eta|<4.5$), corresponds to the $\ZeroJet$ fraction from the 
normalised differential cross section measured as a function of $\Njet$ (see Table~\ref{tab:njets_norm}).  Results for the jet selection in this analysis, and thresholds of
$30\GeV$ and $40\GeV$, are given in Table~\ref{tab:JVE} and compared to predictions in Figure~\ref{fig:JVE}.
The predictions are calculated with JetVHeto, ST, STWZ, N$^{3}$LO+NNLL+LL\_R, and {\POWHEG} {\NNLOPS}, as described in Section~\ref{sec:theopred}.
The results are in agreement with the predictions.  
The predictions are more precise than the measurements reported here, which are limited by their large statistical uncertainties.

\begin{table}[tbp!]
\centering
\caption{\label{tab:JVE} Measured and predicted jet-veto efficiency $\varepsilon_{0}$ for different jet $\pT$ thresholds and the associated statistical and systematic uncertainties. The asterisk for the $25\GeV$ column header
  indicates that the results are for a mixed $\pT$ threshold, which is raised from $25\GeV$ to $30\GeV$ for jets with $2.4<|\eta|<4.5$,
	corresponding to the selection used to define the signal regions for the analysis. Total uncertainties in the measurement are
  given along with their relative composition in terms of source.}  
\begin{tabular}{lrrr} 
\dbline
Jet $\pT$ threshold & $25\GeV$* & $30\GeV$ & $40\GeV$ \\ 
\sgline
$\varepsilon_{0}$ & 0.53 & 0.57 & 0.64\\
Statistical uncertainty  & 0.11 & 0.11 & 0.12 \\
Total uncertainty    & 0.14 & 0.14 & 0.17 \\
\sgline
Predicted $\varepsilon_{0}$ (NNLOPS)	& 0.59	& 0.63	& 0.73	\\
Uncertainty in prediction		& 0.04	& 0.04	& 0.04	\\
\sgline
SR data statistical  &19\% & 17\% & 17\% \\
MC statistical  & 4\% & 3\% & 3\% \\
CR data statistical  & 9\% & 7\% & 8\% \\
Exp. JER & 0\% & 2\% & 3\% \\
Exp. JES & 3\% & 3\% & 5\% \\
Exp. $b$-tag  & 3\% & 3\% & 4\% \\
Exp. leptons  &  2\% & 2\% & 2\% \\
Exp. $\MET$  &  1\% & 1\% & 1\% \\
Exp. other  &  2\% & 2\% & 5\% \\
Theory ($WW$) & 12\% & 11\% & 12\% \\
Theory (top)  & 7\% & 7\% & 9\% \\
Theory (other backgrounds) & 6\% & 5\% & 8\% \\
Theory (signal)   & 1\% & 2\% & 2\% \\
Detector corrections  & <1\% & <1\% & <1\% \\
\sgline
Total &26\% & 24\% & 27\% \\
\dbline
\end{tabular}
\end{table}

\begin{figure}[!tbh]
  \centering
    \includegraphics[width=0.46\textwidth]{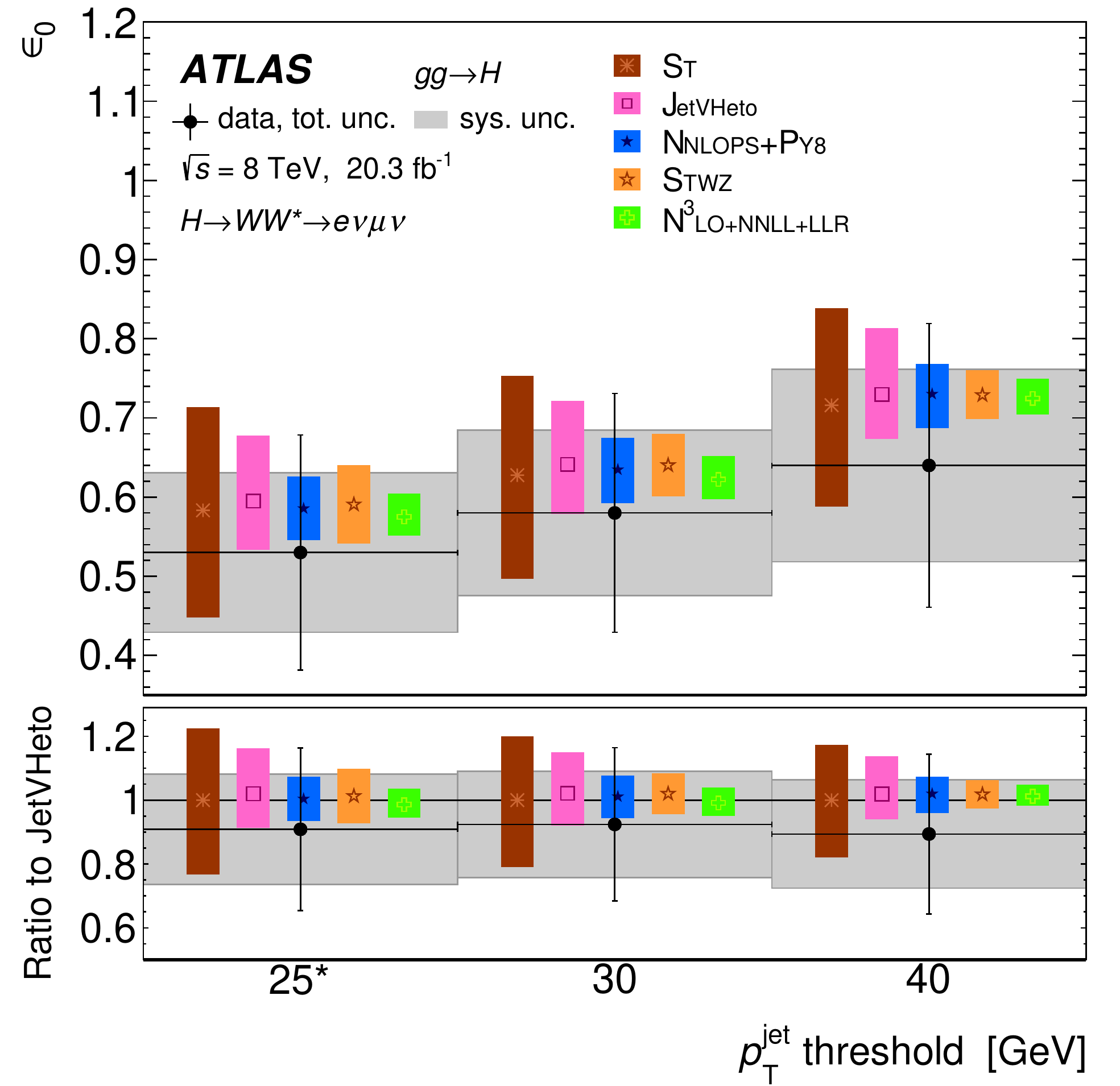}
  { \caption{Measured jet-veto efficiency as a function of the jet $\pT$ threshold, compared to the signal predictions. 
  The asterisk on the $25\GeV$ bin label indicates that the results are for a mixed $\pT$ threshold, which is raised from 
    $25\GeV$ to $30\GeV$ for jets with $2.4<|\eta|<4.5$, corresponding to the selection used to define the signal regions for the analysis.
	     The total uncertainty includes all statistical, experimental, and theoretical uncertainties. 
  \label{fig:JVE}
    }}
\end{figure}

\FloatBarrier

\section{Conclusion}
\label{sec:conclusion}
Measurements of total and differential fiducial cross sections in the \gghwwenmn\  final state of gluon-fusion Higgs boson production are presented.
They are based on $20.3\,\ifb$ of proton--proton 
collision data produced at a centre-of-mass energy of $\rts = 8 \tev$ at the LHC and recorded by the ATLAS experiment in 2012. The data are corrected for detector efficiencies and resolution using an iterative Bayesian method. Results are presented in a fiducial region requiring two opposite-charge leptons of different flavour and missing transverse momentum of more than $20 \gev$. Additional selection requirements are applied on the dilepton system to select Higgs boson candidate events.
The fiducial cross section of ggF Higgs boson production is measured to be:
\begin{linenomath}
\begin{equation}
\sigma_{\mathrm{ggF}}^{\mathrm{fid}} = 36.0 \pm 7.2\mathrm{(stat)} \pm 6.4\mathrm{(sys)} \pm 1.0\mathrm{(lumi)}\; \mathrm{fb}
\end{equation}
\end{linenomath}
for a Higgs boson of mass $125.0 \gev$ produced in the fiducial region described in Table~\ref{tab:fidsel}. 
The SM prediction is $\sigma_{\mathrm{ggF}}^{\mathrm{fid}} = 25.1 \pm 2.6\; \mathrm{fb}$. 

In addition, differential and normalised differential cross sections are measured in the fiducial region as functions of the number of jets, the Higgs boson transverse momentum, the rapidity of the dilepton system, and the 
transverse momentum of the leading jet. These measurements probe directly the Higgs boson production and decay kinematics, as well as the jet activity produced in association with 
the Higgs boson. Jet-veto efficiencies for $H+0$-jet events are also reported for three different thresholds for the transverse momentum 
of the leading jet; the jet-veto efficiency for a threshold of $30 \GeV$ is $(57 \pm 14)$\%. All results are 
compared to a set of predictions from fixed-order calculations and Monte-Carlo generators and are in agreement with the predictions of the Standard Model.

\section*{Acknowledgements}

We thank CERN for the very successful operation of the LHC, as well as the
support staff from our institutions without whom ATLAS could not be
operated efficiently.

We acknowledge the support of ANPCyT, Argentina; YerPhI, Armenia; ARC, Australia; BMWFW and FWF, Austria; ANAS, Azerbaijan; SSTC, Belarus; CNPq and FAPESP, Brazil; NSERC, NRC and CFI, Canada; CERN; CONICYT, Chile; CAS, MOST and NSFC, China; COLCIENCIAS, Colombia; MSMT CR, MPO CR and VSC CR, Czech Republic; DNRF and DNSRC, Denmark; IN2P3-CNRS, CEA-DSM/IRFU, France; GNSF, Georgia; BMBF, HGF, and MPG, Germany; GSRT, Greece; RGC, Hong Kong SAR, China; ISF, I-CORE and Benoziyo Center, Israel; INFN, Italy; MEXT and JSPS, Japan; CNRST, Morocco; FOM and NWO, Netherlands; RCN, Norway; MNiSW and NCN, Poland; FCT, Portugal; MNE/IFA, Romania; MES of Russia and NRC KI, Russian Federation; JINR; MESTD, Serbia; MSSR, Slovakia; ARRS and MIZ\v{S}, Slovenia; DST/NRF, South Africa; MINECO, Spain; SRC and Wallenberg Foundation, Sweden; SERI, SNSF and Cantons of Bern and Geneva, Switzerland; MOST, Taiwan; TAEK, Turkey; STFC, United Kingdom; DOE and NSF, United States of America. In addition, individual groups and members have received support from BCKDF, the Canada Council, CANARIE, CRC, Compute Canada, FQRNT, and the Ontario Innovation Trust, Canada; EPLANET, ERC, FP7, Horizon 2020 and Marie Sk{\l}odowska-Curie Actions, European Union; Investissements d'Avenir Labex and Idex, ANR, R{\'e}gion Auvergne and Fondation Partager le Savoir, France; DFG and AvH Foundation, Germany; Herakleitos, Thales and Aristeia programmes co-financed by EU-ESF and the Greek NSRF; BSF, GIF and Minerva, Israel; BRF, Norway; Generalitat de Catalunya, Generalitat Valenciana, Spain; the Royal Society and Leverhulme Trust, United Kingdom.

The crucial computing support from all WLCG partners is acknowledged
gratefully, in particular from CERN and the ATLAS Tier-1 facilities at
TRIUMF (Canada), NDGF (Denmark, Norway, Sweden), CC-IN2P3 (France),
KIT/GridKA (Germany), INFN-CNAF (Italy), NL-T1 (Netherlands), PIC (Spain),
ASGC (Taiwan), RAL (UK) and BNL (USA) and in the Tier-2 facilities
worldwide.

\printbibliography

\clearpage


\begin{flushleft}
{\Large The ATLAS Collaboration}

\bigskip

G.~Aad$^\textrm{\scriptsize 86}$,
B.~Abbott$^\textrm{\scriptsize 113}$,
J.~Abdallah$^\textrm{\scriptsize 151}$,
O.~Abdinov$^\textrm{\scriptsize 11}$,
B.~Abeloos$^\textrm{\scriptsize 117}$,
R.~Aben$^\textrm{\scriptsize 107}$,
M.~Abolins$^\textrm{\scriptsize 91}$,
O.S.~AbouZeid$^\textrm{\scriptsize 137}$,
H.~Abramowicz$^\textrm{\scriptsize 153}$,
H.~Abreu$^\textrm{\scriptsize 152}$
% ...
\bigskip
\\
% ...
$^{11}$ Institute of Physics, Azerbaijan Academy of Sciences, Baku, Azerbaijan\\
% ...
$^{86}$ CPPM, Aix-Marseille Universit{\'e} and CNRS/IN2P3, Marseille, France\\
% ...
$^{91}$ Department of Physics and Astronomy, Michigan State University, East Lansing MI, United States of America\\
% ...
$^{107}$ Nikhef National Institute for Subatomic Physics and University of Amsterdam, Amsterdam, Netherlands\\
% ...
$^{113}$ Homer L. Dodge Department of Physics and Astronomy, University of Oklahoma, Norman OK, United States of America\\
% ...
$^{117}$ LAL, Univ. Paris-Sud, CNRS/IN2P3, Universit{\'e} Paris-Saclay, Orsay, France\\
% ...
$^{137}$ Santa Cruz Institute for Particle Physics, University of California Santa Cruz, Santa Cruz CA, United States of America\\
% ...
$^{151}$ Institute of Physics, Academia Sinica, Taipei, Taiwan\\
$^{152}$ Department of Physics, Technion: Israel Institute of Technology, Haifa, Israel\\
$^{153}$ Raymond and Beverly Sackler School of Physics and Astronomy, Tel Aviv University, Tel Aviv, Israel\\
% ...
\end{flushleft}


\end{document}